\documentclass[letterpaper]{article}
\pdfoutput=1

\usepackage{
    amssymb, 
    authblk, 
    bm, 
	cite, 
    graphicx, 
    mathtools, 
	xcolor, 
    xspace 
}

\usepackage[margin=1in]{geometry}

\usepackage[pagebackref]{hyperref}

\usepackage[utf8]{inputenc}

\usepackage[nameinlink,capitalize]{cleveref}

\crefname{figure}{Figure}{Figures}

\def\KB{{\overline{K}_B}} 
\DeclareMathOperator{\Pic}{Pic} 
\newcommand*{\matSpace}[3]{M_{#1 \times #2}(#3)} 
\newcommand*{\PD}[1]{[#1]}
\DeclareMathOperator{\Span}{span} 
\DeclarePairedDelimiter\abs{\lvert}{\rvert}
\newcommand\OSCAR{\texttt{OSCAR}\xspace} 
\newcommand\FTheoryTools{\texttt{FTheoryTools}\xspace} 
\newcommand{\OO}{\mathcal{O}} 
\DeclareMathOperator{\GL}{GL} 
\DeclareMathOperator{\U}{U} 
\DeclareMathOperator{\SU}{SU} 
\newcommand*{\asu}{\mathfrak{su}} 
\newcommand*{\asp}{\mathfrak{sp}} 
\newcommand*{\au}{\mathfrak{u}} 
\newcommand\cL{{\mathcal L}} 
\newcommand\QQ{\mathbb{Q}} 
\newcommand\PP{\mathbb{P}} 

\definecolor{bg}{rgb}{0.95,0.95,0.95} 
\usepackage[frozencache,cachedir=./my-minted]{minted}
\hypersetup{
	pdftitle={FTheoryToolsPaper},
	pdfauthor={Martin Bies, Mi{\c k}elis E. Mi{\c k}elsons, Andrew P. Turner},
	pdfsubject={String theory, F-theory, elliptic fibration, Weierstrass model, Tate model, Kodaira classification, resolution of singularities, (toric) blow-ups, intersection theory, Mordell--Weil group, Chow ring, literature model database}
}

\title{\Huge \FTheoryTools: Advancing Computational Capabilities for F-Theory Research}
\author[1]{\Large Martin Bies\thanks{martin.bies@rptu.de}}
\author[1]{\Large Mi{\c k}elis E. Mi{\c k}elsons\thanks{mikelis.emils.mikelsons@gmail.com}}
\author[2]{\Large Andrew P. Turner \thanks{apturner@vt.edu}}
\affil[1]{\normalsize \emph{Department of Mathematics, RPTU Kaiserlautern-Landau, Kaiserslautern, Germany}\vspace*{1em}}
\affil[2]{\normalsize \emph{Physics Department, Robeson Hall, Virginia Tech, Blacksburg, VA 24061, USA}}
\date{}

\begin{document}

\maketitle

\begin{abstract}
A primary goal of string phenomenology is to identify realistic four-dimensional physics within the landscape of string theory solutions. In F-theory, such solutions are encoded in the geometry of singular elliptic fibrations, whose study often requires particularly challenging and cumbersome computations. In this work, we introduce \FTheoryTools, a novel software module integrated into the \texttt{OSCAR} computer algebra system, designed to automate the complex and tedious tasks involved in F-theory model building. Key features of \FTheoryTools include the enumeration of \(G_4\)-fluxes, the capability to perform blowups on arbitrary (including non-toric) loci, and a literature database of existing F-theory constructions employing a MaRDI-based data format for enhanced collaboration and reproducibility. As a demonstration of its power, we present a stress test by applying \FTheoryTools to the challenging \emph{F-theory geometry with most flux vacua}~\cite{Taylor:2015xtz}. Our results illustrate the potential of \FTheoryTools to streamline F-theory research and pave the way for future developments in the computational study of string phenomenology.
\end{abstract}

\newpage

\tableofcontents

\newpage

\section{Introduction}

String phenomenology seeks to derive realistic four-dimensional physics from string theory by constructing compactification geometries that yield viable effective theories. Extensive research---see, e.g., \cite{Candelas:1985en,Greene:1986ar,Berkooz:1996km,Aldazabal:2000dg,Aldazabal:2000cn,Ibanez:2001nd,Blumenhagen:2001te,Cvetic:2001tj,Cvetic:2001nr,Braun:2005ux,Bouchard:2005ag,Gomez:2005ii,Blumenhagen:2005mu,Bouchard:2006dn,Bouchard:2008bg,Donagi:2008ca,Hayashi:2008ba,Donagi:2009ra,Anderson:2009mh,Heckman:2010bq,Anderson:2011ns,Braun:2011zm,Marsano:2011hv,Anderson:2012yf,Krause:2012yh,Intriligator:2012ue,Braun:2013nqa,Cvetic:2013uta,Cvetic:2015txa,Lin:2015qsa,Lin:2016vus,Jefferson:2021bid,Bies:2023jqg,Li:2023dya,Arena:2023maa} and references therein---has revealed a rich landscape of solutions. 
Within the broader context of string phenomenology, F-theory focuses on solutions encoded in the geometry of singular elliptic fibrations, the study of which requires many intricate calculations such as the crepant resolution of singularities. Motivated by the many recent advancements in data science and symbolic computation, we introduce \FTheoryTools, a submodule of the \texttt{OSCAR} computer algebra system \cite{OscarBook, OSCAR}. \FTheoryTools automates many of the complex, tedious, and time-consuming tasks inherent in F-theory model building, thereby streamlining research and enhancing reproducibility. We highlight a few of its features below.

\paragraph{Enumeration of \(G_4\)-Fluxes}  
A central aspect of F-theory research is the study of \(G_4\)-fluxes. \FTheoryTools can enumerate \(G_4\)-fluxes for a wide range of geometries. As a stress test, we demonstrate its ability to compute the \(G_4\)-fluxes for the notably challenging \emph{F-theory geometry with most flux vacua}~\cite{Taylor:2015xtz}. This computation---impractical to perform by hand---is reduced to an operation executable in mere minutes, leveraging precompiled arithmetic results~\cite{bmtMaRDIFiles}.

\paragraph{Blowups on General Loci}
A significant limitation of many F-theory studies is their reliance on toric geometries. \FTheoryTools allows resolution via blowups on arbitrary (including non-toric) loci, extending the scope of accessible F-theory constructions.

\paragraph{Literature Database of Existing F-Theory Constructions} \FTheoryTools provides a growing database of known F-theory constructions, enabling researchers to recreate celebrated models and perform consistency checks with ease. The database employs a data format developed through the MaRDI framework~\cite{mrdi-file-format} in accordance with FAIR principles (Findability, Accessibility, Interoperability, Reusability). This format is based on JSON, a widely used, human-readable data format, and is designed for compatibility with multiple computer algebra systems. Currently, fully functional save/load cycles are supported only for \texttt{OSCAR}, while other systems such as \texttt{SAGE}~\cite{sagemath} and \texttt{Macaulay2}~\cite{M2} can parse the data with appropriate handling. Looking forward, the goal is for the MaRDI format to become a common standard across computer algebra systems, enabling seamless integration without the need for system-specific data handling. In its current state, the database already supports several families of geometries, such as~\cite{Lawrie:2012gg, KMOPR15}, with additional support for constructions in~\cite{Krause:2011xj, Morrison:2012ei, Cvetic:2015, Taylor:2015xtz, CHLLT19}. We envision that this MaRDI-based data format will streamline collaboration and enhance reproducibility by enabling researchers to save, share, and modify F-theory constructions with minimal effort.

\addtocontents{toc}{\protect\setcounter{tocdepth}{1}}
\subsection{Current Status, Installation, Documentation, and Tutorials}
\addtocontents{toc}{\protect\setcounter{tocdepth}{2}}

\FTheoryTools is currently an experimental module of \OSCAR. While we strive for stability in function names and outputs, further refinements are expected during this developmental phase. This paper presents examples generated using \OSCAR~v1.5.0 and Julia~v1.11.6~\cite{Julia-2017}. Approximate runtimes for various tasks are provided as guidelines, subject to variation depending on hardware. To use \FTheoryTools, please install \OSCAR version~1.5.0 or newer. Installation instructions are available at:
\begin{center}
    \href{https://www.oscar-system.org/install/}{https://www.oscar-system.org/install/}
\end{center}
To benefit from ongoing development, we recommend keeping \OSCAR up to date. Documentation for \FTheoryTools is available at:
\begin{center}
    \href{https://docs.oscar-system.org/stable/Experimental/FTheoryTools/introduction/}{https://docs.oscar-system.org/stable/Experimental/FTheoryTools/introduction/}
\end{center}
The accompanying tutorial, as well as an additional notebook containing all the code snippets presented in this article, is also available on the \OSCAR homepage:
\begin{center}
    \href{https://www.oscar-system.org/tutorials/FTheoryTools/}{https://www.oscar-system.org/tutorials/FTheoryTools/},
    \href{https://github.com/HereAround/MartinsOscarTutorials.jl/blob/master/FTheoryToolsPaper.ipynb}{https://github.com/HereAround/MartinsOscarTutorials.jl/blob/master/FTheoryToolsPaper.ipynb}.
\end{center}

\addtocontents{toc}{\protect\setcounter{tocdepth}{1}}
\subsection{Structure of the Paper and Results}
\addtocontents{toc}{\protect\setcounter{tocdepth}{2}}

In \cref{sec:TecForFTheory}, we give a theoretical overview, reviewing several  commonly employed models of singular elliptic fibrations and the crepant resolution of singularities in F-theory. In \cref{sec:DesignFTheoryModelsInOSCAR}, we mirror these theoretical considerations with the corresponding algorithmic framework implemented in \FTheoryTools, including the construction of models and the classification and resolution of singularities. \cref{sec:LiteratureModels} details our data format and literature database, which we believe will facilitate the sharing, verification, and modification of F-theory constructions with minimal effort. In \cref{sec:EnumerationOfG4Fluxes} we will then discuss the enumeration of $G_4$-fluxes with \FTheoryTools. The general framework and the employed computational techniques will be introduced in \cref{subsec:ComputingG4s}. Subsequently, we present a stress test in \cref{sec:g4StressTest} by applying \FTheoryTools's $G_4$-enumeration techniques to the \emph{F-theory geometry with most flux vacua}~\cite{Taylor:2015xtz}, a singular hypersurface in a 5-dimensional toric ambient space defined by 101 rays and 198 maximal cones, cut out by a hypersurface equation polynomial consisting of 355,785 monomials. We resolve the singularities of the hypersurface by executing a sequence of 206 toric blowups. It proves advantageous for our computational framework to also resolve singularities of the toric ambient space, which requires three additional toric blowups. \FTheoryTools not only computes the resolved geometry, but is also able to find a large space of candidate vertical fluxes preserving the gauge group, isomorphic to \(\mathbb{Z}^{224} \times \mathbb{Q}^{127}\). We conclude with a summary and an outlook on future developments in \cref{sec:Outlook}.

\section{Foundations of F-theory and Elliptic Fibrations} \label{sec:TecForFTheory}

In this section, we introduce the foundational concepts and notation for elliptic fibrations in F-theory, which we will later mirror in \cref{sec:DesignFTheoryModelsInOSCAR} with implementations in the computer algebra system \OSCAR. While the F-theory community is certainly familiar with the topics discussed in this section (such as Weierstrass models, the Kodaira classification of singularities, \ldots), it is crucial to establish a precise notation that aligns with our computational framework.

\subsection{Weierstrass Models} \label{subsec:WeierstrassModels}

Weierstrass models are one of the most common presentations of singular elliptic fibrations. Consider a complex K{\"a}hler manifold (or variety) $B$, which will serve as the base of the fibration. We consider a line bundle $\cL \in \Pic(B)$ and the projective bundle\footnote{We use the classical convention for the projectivization of a vector bundle, where the fibers are the lines passing through the origin rather than the hyperplanes. This is the opposite of the convention in \cite{Hartshorne1977}. Also, note that we reserve the usual projection symbol $\pi$ for the projection of an elliptic fibration, i.e., $\pi\colon Y \twoheadrightarrow B$. To avoid confusion, we thus use the symbol $\rho$ for the projection of the ambient projective bundle.}
\begin{equation}
    \rho\colon \PP^{2, 3, 1}[\cL^2 \oplus \cL^3 \oplus \OO_{B}] \twoheadrightarrow B\,.
\end{equation}
Let $X = \PP^{2, 3, 1}[\cL^2 \oplus \cL^3 \oplus \OO_{B}]$ be the total space of this projective bundle. A section of the bundle $\OO_X(6) \otimes \rho^*\cL^6$ then cuts out an elliptic fibration $Y$ as a hypersurface in $X$. We can describe such a section explicitly. To this end, let us denote the homogeneous coordinates of the ambient $\PP^{2, 3, 1}$-bundle as $[x : y : z]$. We see that $x$, $y$, and $z$ are sections of $\OO_X(2) \otimes \rho^*\cL^2$, $\OO_X(3) \otimes \rho^*\cL^3$, and $\OO_X(1)$, respectively. The elliptic fibration is then realized as the locus cut out by the equation 
\begin{equation}
    y^2 = x^3 + f \cdot x z^4 + g \cdot z^6\,, \label{equ:Weierstrass}
\end{equation}
where $f$ and $g$ are sections of the line bundles $\cL^4$ and $\cL^6$, respectively.\footnote{Here we use a standard abuse of notation, suppressing the fact that $f$ and $g$ must be pulled back to sections of line bundles over the ambient space in defining the Weierstrass equation.} For F-theory constructions, we require the elliptic fibration to be Calabi--Yau. This imposes the condition
\begin{equation}
    c_1(\cL) = c_1(B) = \KB\,,
\end{equation}
with $K_B$ the canonical class of the base, so that $\cL$ is the anticanonical bundle.\footnote{In an abuse of notation, \(\overline{K}_B\) denotes both the anticanonical class and the anticanonical bundle of \(B\).}

\subsection{The Refined Tate Fiber Type}

\begin{figure}[tb]
\centering
\includegraphics[width=\textwidth]{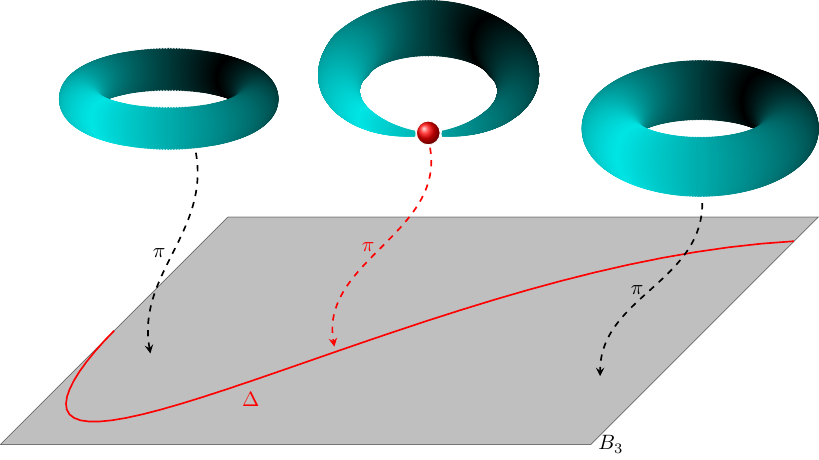}
\caption{A singular elliptic fibration: a smooth elliptic curve over a generic point becomes singular over the discriminant locus \(\mathbb{V}(\Delta)\) (codimension one in \(B\)). Image taken from~\cite{Bies:2018uzw}.}
\label{Figure:F-theorySingularFiberationWithDiscriminant}
\end{figure}

It is well-known (see, e.g.,~\cite{Wei18} for more background) that nontrivial F-theory physics requires a singular Weierstrass model. Therefore, a study of the singularity structure of Weierstrass models is in order. Singular fibers occur over loci of at least complex codimension one in the base $B$, with further singularity enhancements over loci of higher codimension. The singular locus is called the \emph{discriminant locus}, as it is the zero locus $\mathbb{V}(\Delta)$ of the discriminant $\Delta = 4 f^3 + 27 g^2$. An illustrative depiction is provided in~\cref{Figure:F-theorySingularFiberationWithDiscriminant}.

The possible singularity types supported at codimension one have been classified by Kodaira~\cite{Kodaira63,Kodaira63-2} and N{\'e}ron~\cite{Neron64} in what is commonly referred to as the \emph{Kodaira classification}. A codimension-one singularity is characterized by the orders of vanishing of $f$, $g$, and $\Delta$ along the locus. In actuality, the original classification by Kodaira and N{\'e}ron considered only elliptically fibered surfaces. In the F-theory context, where the base may have complex dimension greater than one, an additional feature arises: Tate monodromy. The full data, including both the vanishing orders and the Tate monodromy, specifies the \emph{refined Tate fiber type}. To each such fiber type, one associates a compact simple Lie algebra, which in the physical theory becomes a summand of the gauge algebra.

We will delay the discussion of higher-codimension singularity enhancements and their physical relevance to \cref{sec:resolved-geometry}.

\begin{figure}[tb]
\includegraphics[width=\textwidth]{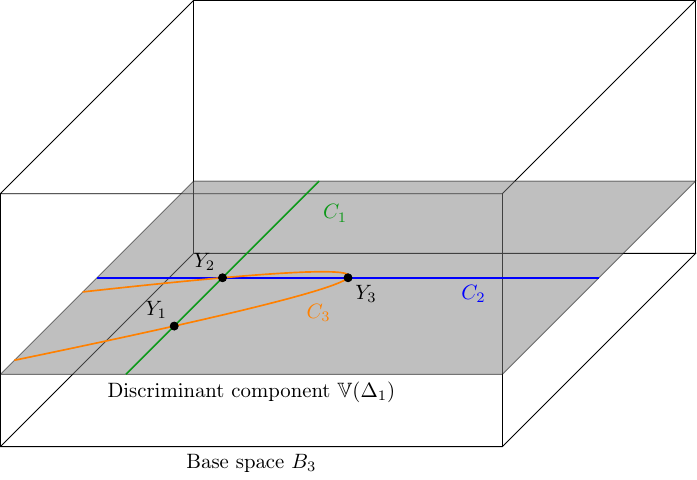}
\caption{An irreducible component \(\mathbb{V}(\Delta_1)\) of the discriminant locus \(\mathbb{V}(\Delta_0\Delta_1\dotsm)\) in a three complex-dimensional base \(B\), showing matter curves \(C_1\), \(C_2\), and \(C_3\) whose intersections yield Yukawa points \(Y_1\), \(Y_2\), and \(Y_3\). Image modified from~\cite{Bies:2018uzw}.}
\label{Figure:F-theoryManyMatterCurvesAndYukawaPoints}
\end{figure}

\subsection{Global Tate Models}

Given the importance of the fiber singularities over the discriminant locus $\mathbb{V}(\Delta)$, one often wishes to engineer fibers of a prescribed refined Tate fiber type. A systematic approach is provided by Tate's algorithm \cite{Tate1975}.

In a sufficiently small open neighborhood $U \subset B$ of a point $p \in B$, the singular elliptic fiber can be described as the hypersurface in $\PP^{2,3,1}$ defined by the polynomial~\cite{Bershadsky:1996nh}
\begin{equation}
    y^2 + a_1(q) x y z + a_3(q) y z^3 = x^3 + a_2(q) x^2 z^2 + a_4(q) x z^4 + a_6(q) z^6\,, \label{equ:TatePolynomial}
\end{equation}
where the coefficients $a_i$ are (local) sections of $\KB^{\otimes i}$ over $U$. This formulation is known as a \emph{Tate model}. Much like the Kodaira classification, in a Tate model the singularities of the elliptic fiber are characterized by the vanishing orders of the coefficients $a_i$, although in this local setting no additional monodromy needs to be taken into account (with a few notable exceptions). The complete classification is summarized in the Tate table; see, e.g.,~\cite{Wei18, Huang:2018gpl}.

In practice, engineering the desired fiber singularities over $\mathbb{V}(\Delta)$ is often simpler in the Tate form than in the Weierstrass form. One may then define the elliptic fibration globally by specifying a single Tate polynomial as in \cref{equ:TatePolynomial}. Such constructions, known as \emph{global Tate models}, are strictly less general than Weierstrass models but have proven very useful for model building in F-theory. A concrete instance, which we discuss in \cref{subsec:ConstructTateModels} with the help of computations in \OSCAR, is found in \cite{Krause:2011xj}.

\subsection{Other Model Presentations} \label{subsec:OtherPresentations}

There exist many other presentations of elliptic fibrations in the literature, several of which have been implemented or are planned for implementation in \FTheoryTools; we mention only a few here.

Both Weierstrass models and global Tate models are examples of \emph{hypersurface models}: descriptions in which the elliptic fibration is defined as the vanishing locus of a single polynomial in an ambient space. Although every model is birationally equivalent to a Weierstrass model (in characteristics other than 2 or 3), for practical purposes it is often more convenient to work with an alternative hypersurface description. An example of such models is given in~\cite{KMOPR15}. We will discuss this important class of F-theory models and the available \FTheoryTools infrastructure in \cref{subsec:FTheoryOnAllToricHypersurfaces}.

Hypersurface models are special cases of \emph{complete intersection models}, where the elliptic fibration is realized as a codimension-$m$ locus in an ambient space defined by the simultaneous vanishing of $m$ polynomials. Complete intersection models are not yet supported by \FTheoryTools.

\subsection{Geometry of the Resolved Elliptic Fibration} \label{sec:resolved-geometry}

While the singularities of the elliptic fibration are crucial for nontrivial F-theory physics, many physical quantities are most easily extracted from a smooth geometry. In practice, one usually proceeds by resolving the singularities---at least as far as possible---to work with a smooth (or nearly smooth) space. For our purposes, we assume the existence of a smooth, flat, and crepant resolution $\widehat{\pi}\colon \widehat{Y} \twoheadrightarrow B$ of the singular elliptic fibration $Y$. To extract the physical quantities of $Y$ from $\widehat{Y}$, we must have $c_1(\widehat{Y}) = 0$, which is why we focus exclusively on crepant resolutions.

It is important to note that crepant resolutions are not natural from the perspective of algebraic geometry. As a result, crepant resolutions are much less studied than non-crepant resolutions. In particular, no algorithm for general crepant resolutions is known as of this writing. Furthermore, it is well-known that $\QQ$-factorial terminal singularities cannot be resolved crepantly. A further challenge arises from the resolution of non-minimal singularities. In this case, one encounters non-flat fibrations, which means that the dimension of the fiber increases over certain loci, see, e.g., \cite{Apruzzi:2018nre}. A typical example is that over certain loci of $B$ the fiber is a complex surface, whereas for generic loci of $B$ the fiber is the union of a collection of curves.

Unless stated differently, in this article we assume that there are neither $\QQ$-factorial terminal singularities nor non-minimal singularities. Then, the primary difference between the singular fibration $Y$ and its (smooth, flat, crepant) resolution $\widehat{Y}$ lies in the structure of the fiber over the discriminant locus $\mathbb{V}(\Delta)$, where the original elliptic fibers are singular. Over a generic point of an irreducible component of $\mathbb{V}(\Delta)$, the resolved fiber splits into a collection of rational curves whose intersection pattern forms a (simply laced) affine Dynkin diagram. In the absence of monodromy (i.e., with trivial Tate monodromy), this diagram is precisely the one associated with the refined Tate fiber type of the singularity. In cases with nontrivial Tate monodromy, the local fiber structure gives rise to a folded version of the Dynkin diagram. In either situation, the Lie algebra corresponding to the (folded) Dynkin diagram appears as a summand in the gauge algebra of the physical theory.

As mentioned previously, higher-codimension singularity enhancements also play a crucial role in F-theory model building. Charged matter typically localizes along codimension-two singular loci (referred to as \emph{matter curves} \(C_{\bm{R}} \subset B\) in the physically relevant case $\dim(B) = 3$) where the fiber singularity enhances. The structure of the generic fiber over such loci fixes the representation \(\bm{R}\) of the gauge algebra in which this matter transforms~\cite{Katz:1996xe}. Codimension-three loci where the singularity type enhances further (known as \emph{Yukawa points} when $\dim(B) = 3$) lead to interaction terms among the charged matter fields, the details of which are again determined by the fiber structure.

\begin{figure}[tb]
\centering
\includegraphics[width=0.9\textwidth]{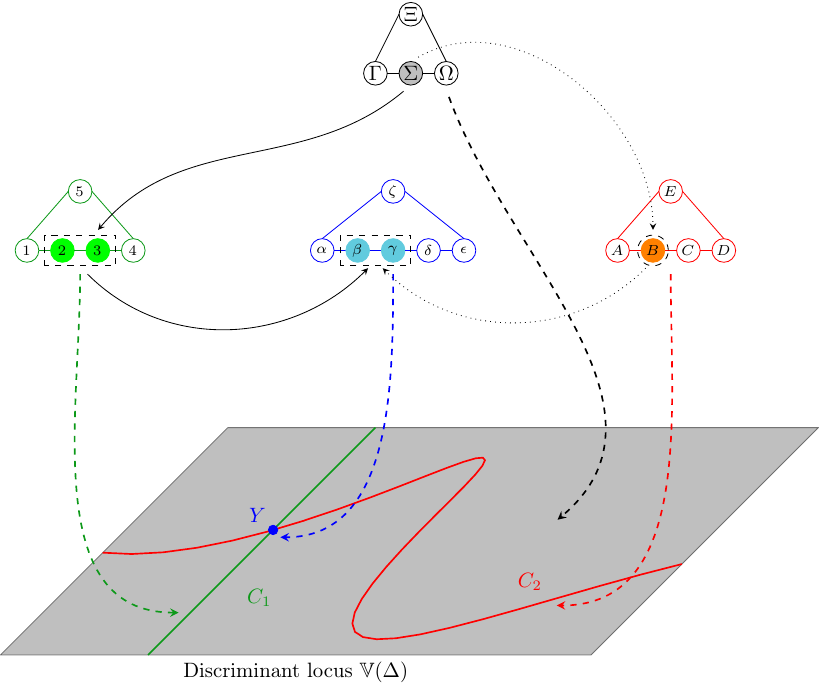}
\caption{A schematic of a resolved elliptic fibration: over a generic point of the discriminant locus \(\mathbb{V}(\Delta)\), the fiber is a set of rational curves forming an affine Dynkin diagram (grey). Over codimension-one and two loci of the discrimant, extra \(\PP^1\)s appear (green, red and blue, respectively). The image tracks the fate of \(\PP^1\) components as we move to higher and higher codimension loci of \(\mathbb{V}(\Delta)\). Image modified from~\cite{Bies:2018uzw}.}
\label{Figure:F-theoryResolvedFiberationOverDiscriminant}
\end{figure}

In the resolved fibration \(\widehat{Y}\), these singularity enhancements manifest as additional \(\PP^1\) components appearing in the fiber. Over a generic codimension-one locus, the fiber consists of rational curves arranged in an affine Dynkin diagram (shown in grey in \cref{Figure:F-theoryResolvedFiberationOverDiscriminant}). Over codimension-two loci, these generic \(\PP^1\) components may split into extra components (illustrated in green and red), and these components may in turn split further at codimension-three loci (illustrated in blue).

\subsection{\texorpdfstring{$G_4$}{G4}-Fluxes}

Let us focus our attention on the physically relevant case of $\dim(B) = 3$. Here, $G_4$-fluxes play a pivotal role in F-theory phenomenology as they determine both the chiral and vector-like matter (see, for instance,~\cite{Bies:2023jqg, Li:2023dya}). A flux $G_4 \in H^{2, 2}(\widehat{Y}_4, \mathbb{R})$
is subject to the quantization condition~\cite{Witten:1996md}:
\begin{equation}
    G_4 + \frac{1}{2} c_2(\widehat{Y}_4) \in H^{4}(\widehat{Y}_4, \mathbb{Z}) \,.
    \label{eq:M2quantization2}
\end{equation}
In many geometries of interest, the second Chern class $c_2(\widehat{Y}_4)$ is even, which simplifies the quantization condition to $G_4 \in H^4_{\mathbb{Z}}(\widehat{Y}_4, \mathbb{Z})$. In addition to the quantization condition \labelcref{eq:M2quantization2}, there are additional conditions that a $G_4$-flux must satisfy in order to not spoil Poincar{\'e} invariance and to preserve minimal supersymmetry; we will discuss these conditions further in \cref{sec:EnumerationOfG4Fluxes}.

Recall from the previous section that---under the assumption of absence of $\QQ$-factorial terminal singularities as well as non-minimal singularities---over a matter curve $C_{\bm{R}} \subseteq {B}$, the reducible fibers of $\widehat{Y}$ consist of a union of rational curves. A state with weight $\mathbf{w}$ in the representation $\bm{R}$ corresponds to a linear combination of these rational curves. Fibering this linear combination over the matter curve $C_{\bm{R}}$ produces the \emph{matter surface} $S^{\mathbf{w}}_{\bm{R}}$. The chiral index of the massless matter localized on said matter curve is then given by the integral
\begin{equation}
\chi \left( \bm{R} \right) = \int_{S^{\mathbf{w}}_{\bm{R}}} G_4 \,,
\end{equation}
as discussed in, e.g., \cite{Donagi:2008ca,Hayashi:2008ba,Donagi:2009ra,Heckman:2010bq,Marsano:2011hv,Grimm:2011fx,Braun:2011zm,Krause:2011xj,Krause:2012yh,Intriligator:2012ue}.

Note that a $G_4$-flux can induce different chiral indices and vector-like spectra for different weight states $\mathbf{w}$, with corresponding matter surfaces $S^{\mathbf{w}}_{\bm{R}}$. In this case, part of the gauge algebra---determined by the codimension-1 fiber singularities---is broken by the given $G_4$-flux. Such breakings are instrumental to F-theory model building. More generally, a typical task in F-theory model building concerns the identification of vertical $G_4$-fluxes with non-negative D3-tadpole that leave a chosen subalgebra of the geometric gauge algebra unbroken. While of central importance, such enumerations are typically technical and tedious. Steps to automate the involved computation were for instance taken in~\cite{Lin:2016vus}, see also references therein. Our intention is to take this technology yet another step further. We will discuss our approach in~\cref{sec:EnumerationOfG4Fluxes}.

\section{Algorithmic Framework for F-theory Models in \OSCAR}
\label{sec:DesignFTheoryModelsInOSCAR}

In this section, we translate the theoretical framework of \cref{sec:TecForFTheory} into explicit computational models using \OSCAR. While the abstract structures of F-theory are well understood, their realization in a computer algebra system introduces subtleties stemming from algorithmic constraints, representation choices, and implementation details. Our goal is to develop a workflow that faithfully reflects the geometry of F-theory while leveraging the computational power of \OSCAR. To this end, we will now present concrete implementations of Weierstrass, Tate, and hypersurface models, along with examples that illustrate how theoretical formulations are adapted for efficient computation. The implementation details of literature models and \(G_4\)-fluxes are discussed in~\cref{sec:LiteratureModels} and~\cref{sec:EnumerationOfG4Fluxes}, respectively.

It is worth noting that working with families of models, defined over arbitrary base spaces, is a common practice in the F-theory community. However, while conceptually standard among researchers, this approach presents additional challenges when implemented in a computer algebra system. For instance: what precisely constitutes the defining data of a family of models? As a result, the current functionality for handling such families of elliptic fibrations in \FTheoryTools is fairly limited. In light of these limitations and the ongoing development efforts, we will not pursue models over families of base spaces further in this article. For more information on this topic and details on the latest implementation, we refer the interested reader to the \OSCAR documentation.

When considering F-theory constructions over a fixed base, our framework supports both toric varieties and more general schemes as base spaces. Most of the functionality in \FTheoryTools is optimized for complete toric varieties. Among other applications, this enables us to easily verify that a given polynomial defines a (local) section of a specified line bundle, and to integrate cohomology classes.

\subsection{Weierstrass Models} \label{subsec:ConstructWeierstrassModels}

As explained in the theoretical discussion, constructing a Weierstrass model requires the specification of a base space and the corresponding Weierstrass sections \(f\) and \(g\), which are sections of line bundles over the base space $B$.

As elaborated above, most functionality is available when the base is a complete toric variety---complete varieties permit computations such as the integration of cohomology classes and the evaluation of line bundle cohomologies, while toric varieties typically provide a computationally favorable setting compared to  schemes. Thus, as an example, let us begin by constructing a Weierstrass model over the base \(B = \mathbb{P}^2\). We denote the homogeneous coordinates of \(\mathbb{P}^2\) by \([x_1 : x_2 : x_3]\). In this case, \(f\) and \(g\) are taken to be sections of the line bundles \(\mathcal{O}_{\mathbb{P}^2}(12)\) and \(\mathcal{O}_{\mathbb{P}^2}(18)\), respectively. For example, we can make the following (somewhat arbitrary) choice:
\begin{equation}
\begin{aligned}
    f &= \frac{1}{48} \left(-\left(28 x_1 x_2^5 + 169 x_3^6\right)^2 + 24 \left(2 x_1^3 (x_2 + x_3)^9 + 13 x_1^2 x_2^4 x_3^6\right)\right)\,, \\
    g &= \frac{1}{864} \left(216 x_1^4 x_2^8 x_3^6 + \left(28 x_1 x_2^5 + 169 x_3^6\right)^3 - 
   36 \left(28 x_1 x_2^5 + 169 x_3^6\right) \left(2 x_1^3 (x_2 + x_3)^9 + 13 x_1^2 x_2^4 x_3^6\right)\right)\,.
\end{aligned}
\end{equation}

The corresponding \OSCAR code is as follows:
\begin{minted}[bgcolor=bg,linenos,breaklines]{julia}
julia> B2 = projective_space(NormalToricVariety, 2)
Normal toric variety

julia> x1, x2, x3 = gens(cox_ring(B2))
3-element Vector{MPolyDecRingElem{QQFieldElem, QQMPolyRingElem}}:
 x1
 x2
 x3

julia> weier_f = 1//48*(-(28*x1*x2^5 + 169*x3^6)^2 + 24*(2*x1^3*(x2 + x3)^9 + 13*x1^2*x2^4*x3^6));

julia> weier_g = 1//864*(216*x1^4*x2^8*x3^6 + (28*x1*x2^5 + 169*x3^6)^3 - 36*(28*x1*x2^5 + 169*x3^6)*(2*x1^3*(x2 + x3)^9 + 13*x1^2*x2^4*x3^6));

julia> w = weierstrass_model(B2, weier_f, weier_g; completeness_check = false)
Weierstrass model over a concrete base

julia> explicit_model_sections(w)
Dict{String, MPolyDecRingElem{QQFieldElem, QQMPolyRingElem}} with 2 entries:
  "f" => x1^3*x2^9 + 9*x1^3*x2^8*x3 + 36*x1^3*x2^7*x3^2 + 84*x1^3*x2^6*x3^3 + 126*x1^3*x2^5*x3^4 + 126*x1^3*x2^4*x3…
  "g" => -7//3*x1^4*x2^14 - 21*x1^4*x2^13*x3 - 84*x1^4*x2^12*x3^2 - 196*x1^4*x2^11*x3^3 - 294*x1^4*x2^10*x3^4 - 294…
\end{minted}
Notice the optional argument \verb|completeness_check| in the constructor. Sanity checks---for instance verifying that the given polynomials for the Weierstrass sections $f$ and $g$ are sections of the correct line bundles---are only available for complete toric bases. As these checks can be computationally demanding, we allow users to disable them as shown above.

Similar to the above, \OSCAR can also generate the most generic Weierstrass model over a given base. To this end, pseudo-random Weierstrass sections $f$ and $g$ are automatically generated:
\begin{minted}[bgcolor=bg,linenos,breaklines]{julia}
julia> B2 = projective_space(NormalToricVariety, 2)
Normal toric variety

julia> w_generic = weierstrass_model(B2; completeness_check = false)
Weierstrass model over a concrete base

julia> explicit_model_sections(w_generic)
Dict{String, MPolyDecRingElem{QQFieldElem, QQMPolyRingElem}} with 2 entries:
  "f" => 5222*x1^12 - 5191*x1^11*x2 + 6322*x1^11*x3 + 8628*x1^10*x2^2 + 639*x1^10*x2*x3 + 9202*x1^10*x3^2 - 2993*x1^9*x2^3 + 8810*x1^9*x2^2*x3 + 1794*x1^9*x2*x3^2 - 7766*x1^9*x3^3 - 1502*x1^8*x2^…
  "g" => 9377*x1^18 - 4524*x1^17*x2 - 7759*x1^17*x3 + 8902*x1^16*x2^2 + 4313*x1^16*x2*x3 - 7190*x1^16*x3^2 + 6192*x1^15*x2^3 + 2846*x1^15*x2^2*x3 - 9632*x1^15*x2*x3^2 - 3004*x1^15*x3^3 + 3145*x1^…
\end{minted}

\subsection{The Refined Tate Fiber Type} \label{subsec:RefinedTateFiberType}

To identify the non-abelian gauge group factors in an F-theory model, one must determine the refined Tate fiber type along codimension-one singular loci. In full generality, this requires tracking the components of the resolved fiber as one traverses a closed loop in the singular locus. However, Tate's algorithm~\cite{Tate1975} shows that the necessary monodromy data can be extracted directly from the singular Weierstrass model by checking a global factorization condition on a \emph{monodromy cover}. Operationally, this amounts to verifying the factorization of an auxiliary polynomial constructed from the Weierstrass coefficients \(f\) and \(g\) evaluated along the singular locus (see~\cite{Grassi:2011hq, Wei18} for details).

Because the local expressions for \(f\) and \(g\) are often very complicated polynomials in the base coordinates, the required Gr{\"o}bner basis calculations can be computationally expensive. To address this, we implement a probabilistic algorithm that reduces the complexity of the computation. For a given codimension-one singular locus defined by \(\sigma = 0\), we randomly reduce \(\sigma\), \(f\), \(g\), and \(\Delta\) to polynomials in only two variables by assigning random integer values to all but two of the local base coordinates. To mitigate the risk of introducing accidental structure, we perform five independent random reductions and compare the outcomes.

Using these reduced polynomials, we construct the polynomial ideals corresponding to the coefficients of the auxiliary polynomial defining the monodromy cover. For all Kodaira types \emph{except} \(I_0^*\), the factorization condition reduces to checking whether a particular coefficient is a perfect square. In practice, this can be effectively determined by checking that the corresponding ideal is radical\footnote{Technically, this does allow for the possibility that the coefficient is a perfect odd power, but this occurrence is rare.}. This final Gr{\"o}bner basis computation typically completes in a reasonable time when using the reduced polynomials.

For \(I_0^*\) singularities, where the monodromy cover is cubic, factorization cannot be reduced to simple conditions on the coefficients. In these cases, we resort to explicit polynomial long division to construct the monodromy cover and then check its factorization directly. This method sometimes fails to properly distinguish split or semi-split \(I_0^*\) singularities due to the non-uniqueness of division in multivariate polynomial rings, which is clearly undesirable. Future improvements will aim to enhance the reliability of the algorithm for \(I_0^*\) singularities as well as the overall probabilistic approach.

To exemplify our algorithm, consider the Weierstrass models over \(B = \mathbb{P}^2\) discussed previously. For the generic Weierstrass model \verb|w_generic|, which employs pseudo-random sections \(f\) and \(g\), only an \(I_1\) locus is expected:
\begin{minted}[bgcolor=bg,linenos,breaklines]{julia}
julia> singular_loci(w_generic)
1-element Vector{Tuple{MPolyIdeal{<:MPolyRingElem}, Tuple{Int64, Int64, Int64}, String}}:
 (Ideal with 1 generator, (0, 0, 1), "I_1")
\end{minted}
In contrast, the Weierstrass model \verb|w|---in the next section, we will see that it is birationally equivalent to a global \(\U(1)\) restricted \(\SU(5)\) Tate model---exhibits both an \(I_1\) locus and a split \(I_5\) singular locus:
\begin{minted}[bgcolor=bg,linenos,breaklines]{julia}
julia> singular_loci(w)
2-element Vector{Tuple{MPolyIdeal{<:MPolyRingElem}, Tuple{Int64, Int64, Int64}, String}}:
 (Ideal with 1 generator, (0, 0, 1), "I_1")
 (Ideal (x1), (0, 0, 5), "Split I_5")
\end{minted}
Note that this output informs us that the split \(I_5\) locus is given by \(\mathbb{P}^1 \cong \mathbb{V}(x_1) \subset \mathbb{P}^2\).

\subsection{Global Tate Models} \label{subsec:ConstructTateModels}

Much of what was discussed for Weierstrass models carries over to global Tate models. In a global Tate model, the Tate coefficients \(a_i\) are specified as sections over the base \(B\). Here, we illustrate a construction over \(B = \mathbb{P}^2\), whose homogeneous coordinates are denoted by \([x_1 : x_2 : x_3]\):
\begin{minted}[bgcolor=bg,linenos,breaklines]{julia}
julia> B2 = projective_space(NormalToricVariety, 2);

julia> x1, x2, x3 = gens(cox_ring(B2));

julia> a1 = 13 * x3^3;
julia> a2 = 7 * x1 * x2^5;
julia> a3 = x1^2 * x2^4 * x3^3;
julia> a4 = x1^3 * (x2 + x3)^9;
julia> a6 = zero(cox_ring(B2));

julia> t = global_tate_model(B2, [a1, a2, a3, a4, a6])
Global Tate model over a concrete base

julia> hypersurface_equation(t)
x1^3*x2^9*x*z^4 + 9*x1^3*x2^8*x3*x*z^4 + 36*x1^3*x2^7*x3^2*x*z^4 + 84*x1^3*x2^6*x3^3*x*z^4 + 126*x1^3*x2^5*x3^4*x*z^4 + …
\end{minted}
By this construction, the vanishing orders of the Tate coefficients along the locus \(\{x_1 = 0\}\) are
\[
\operatorname{ord}\bigl(a_1, a_2, a_3, a_4, a_6\bigr) = \bigl(0,\, 1,\, 2,\, 3,\, \infty\bigr).
\]
From the Tate table (see, e.g.,~\cite{Bershadsky:1996nh,Wei18, Huang:2018gpl}), we deduce the existence of a split \(I_5\)-singularity along \(\{x_1 = 0\}\).

Recall from \cref{subsec:ConstructWeierstrassModels} that we constructed a Weierstrass model \verb|w| with the following $f$, $g$:
\begin{equation}
\begin{aligned}
    f &= \frac{1}{48} \Bigl(-\bigl(28 x_1 x_2^5 + 169 x_3^6\bigr)^2 + 24\Bigl(2 x_1^3 (x_2 + x_3)^9 + 13 x_1^2 x_2^4 x_3^6\Bigr)\Bigr)\,, \\
    g &= \frac{1}{864} \Bigl(216 x_1^4 x_2^8 x_3^6 + \bigl(28 x_1 x_2^5 + 169 x_3^6\bigr)^3 - 
   36 \bigl(28 x_1 x_2^5 + 169 x_3^6\bigr)\Bigl(2 x_1^3 (x_2 + x_3)^9 + 13 x_1^2 x_2^4 x_3^6\Bigr)\Bigr)\,.
\end{aligned}
\end{equation}
We verified that this Weierstrass model exhibits a split \(I_5\)-singularity along $\{ x_1 = 0 \}$. In fact, this Weierstrass model is birational to the above-constructed global Tate model \verb|t|. To illustrate this, we can ask \texttt{OSCAR} to compute the Weierstrass model corresponding to \verb|t|:
\begin{minted}[bgcolor=bg,linenos,breaklines]{julia}
julia> w = weierstrass_model(t)
Weierstrass model over a concrete base

julia> weierstrass_section_f(w)
x1^3*x2^9 + 9*x1^3*x2^8*x3 + 36*x1^3*x2^7*x3^2 + 84*x1^3*x2^6*x3^3 + 126*x1^3*x2^5*x3^4 + 126*x1^3*x2^4*x3^5 + 84*x1^3*x2^3*x3^6 + …

julia> weierstrass_section_g(w)
-7//3*x1^4*x2^14 - 21*x1^4*x2^13*x3 - 84*x1^4*x2^12*x3^2 - 196*x1^4*x2^11*x3^3 - 294*x1^4*x2^10*x3^4 - 294*x1^4*x2^9*x3^5 - …
\end{minted}
For further details, including a convenient display of the Tate table, please refer to the \OSCAR documentation of global Tate models:
\begin{center}
\href{https://docs.oscar-system.org/stable/Experimental/FTheoryTools/tate/}{\small{\mbox{https://docs.oscar-system.org/stable/Experimental/FTheoryTools/tate/}}}.
\end{center}

\subsection{Hypersurface Models} \label{subsec:ConstructHypersurfaceModels}

Recall from \cref{subsec:OtherPresentations} that Weierstrass and global Tate models are special cases of \emph{hypersurface models}, where the fibration is defined as the vanishing locus of a single polynomial in an ambient space. In our framework for hypersurface models---following the approach of~\cite{KMOPR15}---we assume that the fiber ambient space is toric. As described in \cite{KMOPR15}, we choose to parametrize the fibration by two divisors, \(D_1\) and \(D_2\), so that the first two homogeneous coordinates of the fiber ambient space transform as sections of the associated line bundles, respectively, while all other fiber coordinates transform trivially. Any model with toric fiber ambient space can be cast into this form.

For toric bases \(B\), the full toric ambient space \(A\) is constructed by extending the rays and maximal cones of both \(B\) and the toric fiber \(F\). The elliptic fibration then corresponds to a hypersurface in \(A\) that, to satisfy the Calabi--Yau condition, must be a section of the anticanonical bundle \(\overline{K}_A\). For a complete toric base \(B\), multiple valid toric ambient spaces $A$ do exist. The complete list of these toric ambient spaces $A$ can be obtained from a resource-intensive triangulation computation. \FTheoryTools avoids this demanding triangulation task at the expense of determining a single ambient space $A$.

To illustrate this technology, we reconstruct the Tate model \verb|t| from \cref{subsec:ConstructTateModels} within the hypersurface model framework:
\begin{minted}[bgcolor=bg,linenos,breaklines]{julia}
julia> B2 = projective_space(NormalToricVariety, 2)
Normal toric variety

julia> fiber_ambient_space = weighted_projective_space(NormalToricVariety, [2,3,1])
Normal toric variety

julia> set_coordinate_names(fiber_ambient_space, ["x", "y", "z"])

julia> D1 = 2 * anticanonical_divisor_class(B2)
Divisor class on a normal toric variety

julia> D2 = 3 * anticanonical_divisor_class(B2)
Divisor class on a normal toric variety

julia> amb_ring, (x1, x2, x3, x, y, z) = polynomial_ring(QQ, ["x1", "x2", "x3", "x", "y", "z"])
(Multivariate polynomial ring in 6 variables over QQ, QQMPolyRingElem[x1, x2, x3, x, y, z])

julia> p = x^3 + 7*x1*x2^5*x^2*z^2 + x1^3*(x2 + x3)^9*x*z^4 - y^2 - 13*x3^3*x*y*z - x1^2*x2^4*x3^3*y*z^3

julia> h = hypersurface_model(B2, fiber_ambient_space, [D1, D2], p, completeness_check = false)
Hypersurface model over a concrete base

julia> hypersurface_equation(h)
x1^3*x2^9*x*z^4 + 9*x1^3*x2^8*x3*x*z^4 + 36*x1^3*x2^7*x3^2*x*z^4 + 84*x1^3*x2^6*x3^3*x*z^4 + 126*x1^3*x2^5*x3^4*x*z^4 + …
\end{minted}

It is instructive to compare the two models \verb|h| and \verb|t|:
\begin{minted}[bgcolor=bg,linenos,breaklines]{julia}
julia> tate_polynomial(t)
x1^3*x2^9*x*z^4 + 9*x1^3*x2^8*x3*x*z^4 + 36*x1^3*x2^7*x3^2*x*z^4 + 84*x1^3*x2^6*x3^3*x*z^4 + 126*x1^3*x2^5*x3^4*x*z^4 + …

julia> hypersurface_equation(h)
x1^3*x2^9*x*z^4 + 9*x1^3*x2^8*x3*x*z^4 + 36*x1^3*x2^7*x3^2*x*z^4 + 84*x1^3*x2^6*x3^3*x*z^4 + 126*x1^3*x2^5*x3^4*x*z^4 + …

julia> cox_ring(ambient_space(t))
Multivariate polynomial ring in 6 variables over QQ graded by
  x1 -> [1 0]
  x2 -> [1 0]
  x3 -> [1 0]
  x -> [6 2]
  y -> [9 3]
  z -> [0 1]

julia> cox_ring(ambient_space(h))
Multivariate polynomial ring in 6 variables over QQ graded by
  x1 -> [1 0]
  x2 -> [1 0]
  x3 -> [1 0]
  x -> [6 2]
  y -> [9 3]
  z -> [0 1]
\end{minted}
We will discuss support for the models in~\cite{KMOPR15} via our literature database in~\cref{sec:LiteratureModels}. For other examples and background on the hypersurface model implementation, please consult the \OSCAR documentation:
\begin{center}
\href{https://docs.oscar-system.org/stable/Experimental/FTheoryTools/hypersurface/}{\small{\mbox{https://docs.oscar-system.org/stable/Experimental/FTheoryTools/hypersurface/}}}.
\end{center}

\subsection{Resolution of Singularities---Toric Blowups and Beyond}
\label{sec:resolutions}

Extracting physically relevant information from an F-theory model often requires crepantly resolving singular geometries. However, smooth crepant resolutions do not exist in the presence of $\mathbb{Q}$-factorial terminal singularities. Even when such a resolution exists and is known, computing the strict transform of the model can be intricate. Toric resolutions are frequently favored due to their tractability and structured framework, which facilitate the extraction of physical quantities. Here, we present an example of a toric resolution for the $\asu(5) \oplus \au(1)$ singular global Tate model, originally introduced in~\cite{Krause:2011xj}, using \FTheoryTools. We then extend our discussion to non-toric blowups of said geometry.

\subsubsection{A Toric Blowup Sequence}
\label{sec:ToricBlowup}

Global Tate models are defined by the hypersurface equation:
\begin{equation}
y^2 + a_1 x y z + a_3 y z^3 = x^3 + a_2 x^2 z^2 + a_4 x z^4 + a_6 z^6,
\end{equation}
where $a_i \in H^0(B_3, \overline{K}_{B_3}^{\otimes i})$. Following~\cite{Krause:2011xj}, we specialize the Tate sections $a_i$ to enforce an $\asu(5) \oplus \au(1)$ gauge algebra. This is achieved by selecting a divisor $\mathbb{V}(w) \subset B_3$, where $w$ is a generic section of a chosen line bundle $W \in \Pic(\mathbb{P}^3)$. The $\asu(5)$ singularity is imposed by demanding that $a_1$ through $a_4$ be proportional to certain powers of $w$, as shown below, while setting $a_6 = 0$ ensures the presence of the $\au(1)$ summand:
\begin{equation}
a_1 = a_{1,0}, \quad a_2 = w \cdot a_{2,1}, \quad a_3 = w^2 \cdot a_{3,2}, \quad a_4 = w^3 \cdot a_{4,3}, \quad a_6 = 0.
\end{equation}
To construct a simple model that admits a toric resolution, we could make the following choice:
\begin{minted}[bgcolor=bg,linenos,breaklines]{julia}
julia> B3 = projective_space(NormalToricVariety, 3)
Normal toric variety

julia> cox_ring(B3)
Multivariate polynomial ring in 4 variables over QQ graded by
  x1 -> [1]
  x2 -> [1]
  x3 -> [1]
  x4 -> [1]

julia> W = toric_line_bundle(torusinvariant_prime_divisors(B3)[1])
Toric line bundle on a normal toric variety

julia> w = generic_section(W)
772*x1 - 5333*x2 + 6001*x3 - 5121*x4
\end{minted}
Since the coefficients in $w$ are chosen pseudo-randomly, they will vary between executions. Although this choice is technically toric, our algorithm currently does not recognize it as such. Future improvements aim to address this limitation. Fortunately, after a coordinate redefinition, the generic section can always be rewritten as $w = x_1$. Thus, without loss of generality, we make the following choice:
\begin{minted}[bgcolor=bg,linenos,breaklines]{julia}
julia> w = gens(cox_ring(B3))[1]
x1
\end{minted}
Next, we compute the specialized Tate sections $a_{i,j}$:
\begin{minted}[bgcolor=bg,linenos,breaklines]{julia}
julia> Kbar = anticanonical_bundle(B3)
Toric line bundle on a normal toric variety

julia> a10 = generic_section(Kbar);

julia> a21 = generic_section(Kbar^2*W^(-1));

julia> a32 = generic_section(Kbar^3*W^(-2));

julia> a43 = generic_section(Kbar^4*W^(-3));

julia> a6 = zero(cox_ring(B3));
\end{minted}
%
%
%
%
%

With these data, we construct the global Tate model. Additionally, we compute the singular loci and classify the singularity types:
\begin{minted}[bgcolor=bg,linenos,breaklines]{julia}
julia> t = global_tate_model(B3, [a10, a21 * w, a32 * w^2, a43 * w^3, a6])
Global Tate model over a concrete base

julia> singular_loci(t)
2-element Vector{Tuple{MPolyIdeal{<:MPolyRingElem}, Tuple{Int64, Int64, Int64}, String}}:
 (Ideal with 1 generator, (0, 0, 1), "I_1")
 (Ideal (x1), (0, 0, 5), "Split I_5")
\end{minted}
The last line indicates a split $I_5$ singularity over a specific locus in the base. By construction, this locus must be $\{ w = 0 \}$, which we can see that it is.
Note that the output of \verb|singular_loci()| includes loci with singular fibers that are not singularities of the full elliptic fibration; these are precisely the loci with $I_1$ fibers.

One known resolution for this model consists of five toric blowups, as detailed in~\cite{Krause:2011xj}. To outline this sequence, we first inspect the Cox ring of the toric ambient space associated with the singular model:
\begin{minted}[bgcolor=bg,linenos,breaklines]{julia}
julia> amb = ambient_space(t)
Normal toric variety

julia> cox_ring(amb)
Multivariate polynomial ring in 7 variables over QQ graded by
  x1 -> [1 0]
  x2 -> [1 0]
  x3 -> [1 0]
  x4 -> [1 0]
  x -> [8 2]
  y -> [12 3]
  z -> [0 1]
\end{minted}
With these coordinate assignments, we wish to apply the following sequence of blowups (specifying, for each, the center of the blowup and the name of the introduced coordinate $e$ for which $\mathbb{V}(e)$ is the exceptional locus, consistent with~\cite{Krause:2011xj}):
\begin{enumerate}
 \item Center: $\mathbb{V}(x, y, w)$; Exceptional divisor: $e_1$.
 \item Center: $\mathbb{V}(y, e_1)$; Exceptional divisor: $e_4$.
 \item Center: $\mathbb{V}(x, e_4)$; Exceptional divisor: $e_2$.
 \item Center: $\mathbb{V}(y, e_2)$; Exceptional divisor: $e_3$.
 \item Center: $\mathbb{V}(x, y)$; Exceptional divisor: $s$.
\end{enumerate}
Hence, the first step is to blow up the locus $\mathbb{V}(x, y, w)$ within the toric ambient space of \verb|t|. This is implemented in \OSCAR as follows:
\begin{minted}[bgcolor=bg,linenos,breaklines]{julia}
julia> t1 = blow_up(t, ["x", "y", string(w)]; coordinate_name = "e1")
Partially resolved global Tate model over a concrete base
\end{minted}
Instead of repeating such a command four more times, a more efficient approach is available in \FTheoryTools. Specifically, an internal method allows us to automate the entire blowup sequence. However, before utilizing this functionality, we must ensure that the model \verb|t| knows the resolution in question. As created above, it does not:
\begin{minted}[bgcolor=bg,linenos,breaklines]{julia}
julia> resolutions(t)
ERROR: ArgumentError: No resolutions known for this model
\end{minted}
We address this by explicitly specifying the resolution sequence:
\begin{minted}[bgcolor=bg,linenos,breaklines]{julia}
julia> add_resolution!(t, [["x", "y", "w"], ["y", "e1"], ["x", "e4"], ["y", "e2"], ["x", "y"]], ["e1", "e4", "e2", "e3", "s"])
\end{minted}
Here, the character \verb|"w"| appears as a string. To ensure correct interpretation, we explicitly link it to the section $w$ computed earlier:
\begin{minted}[bgcolor=bg,linenos,breaklines]{julia}
julia> explicit_model_sections(t)["w"] = w
x1
\end{minted}
With this information now in place, we can resolve the global Tate model \verb|t|:
\begin{minted}[bgcolor=bg,linenos,breaklines]{julia}
julia> t_res = resolve(t, 1)
Partially resolved global Tate model over a concrete base
\end{minted}
The function \verb|resolve(t, 1)| takes a positive integer $k$ as its second argument, instructing the software to execute the $k$-th known resolution. In this case, it applies the first (and only) resolution stored in the model.

The reader may have noticed that the code sample above prints \emph{partially resolved}. One might ask: is this not a complete resolution of the model? Indeed, in this case, the applied resolution fully resolves the model. However, the term \emph{partially} reflects the fact that verifying this explicitly is computationally expensive. Rather than incurring this cost, we currently skip this verification step. To avoid making any false claims, we adopt a conservative approach and label the output as \emph{partially resolved}. While this decision is based on technical considerations, future improvements to \FTheoryTools aim to refine this behavior.

Automating the resolution step, as demonstrated above, provides an efficiency gain in F-theory model building, significantly reducing the effort required compared to performing blowups, strict transforms, and related operations manually. For example, we can now directly inspect the strict transform of the Tate polynomial (with the large output mostly suppressed here):
\begin{minted}[bgcolor=bg,linenos,breaklines]{julia}
julia> tate_polynomial(t_res)
-1965*x1^16*x*z^4*e1^15*e4^14*e2^14*e3^13 - 5515*x1^15*x2*x*z^4*e1^14*e4^13*e2^13*e3^12 + …
\end{minted}
Despite these advancements, further refinements remain possible. Since the model above has been analyzed in detail in~\cite{Krause:2011xj}, we can leverage this prior work by pre-storing relevant computations and properties within \FTheoryTools. This would allow the model to be automatically configured with the necessary properties when constructed based on this reference. Such a convenient constructor streamlines the workflow by automating tasks such as computing the Tate sections, defining the resolution sequence, specifying the meaning of $w$, and handling other related operations. We refer to these pre-configured setups as \emph{literature models} and discuss them in detail in \cref{sec:LiteratureModels}.

\subsubsection{A Non-Toric Blowup Sequence}
\label{sec:NonToricBlowup}

While toric resolutions are favored for their technical convenience, non-toric resolutions are equally important conceptually. Some F-theory models require methods beyond toric blowups; see, for instance,~\cite{Raghuram:2019efb,Jefferson:2022yya}. To address this, \FTheoryTools provides functionality for handling non-toric blowups.\footnote{It is worth noting that there are some subvarieties that are toric in actuality, but are not currently recognized as toric by our algorithm. For such loci, the non-toric blowup functionality is used. Future improvements aim to refine the algorithm to allow for blowups on arbitrary toric loci using the toric methods.}

Non-toric blowups pose two main challenges: increased computational demand and a different mode of presentation. In \OSCAR, such blowups are handled within the covered scheme framework, requiring explicit computation and evaluation of transition maps, which leads to higher computational cost. Moreover, covered schemes are represented by their affine charts and transition maps. Whereas toric resolutions are often presented in a global description, describing the strict transform after a sequence of non-toric blowups requires the use of affine charts.

To demonstrate this in detail, we return to the model originally introduced in~\cite{Krause:2011xj} and discussed in the previous subsection. We will modify this model so that its resolution requires the use of blowups along non-toric loci. The key modification lies in the choice of the divisor class $W$, which we now adjust as follows:
\begin{minted}[bgcolor=bg,linenos,breaklines]{julia}
julia> W = toric_line_bundle(2 * torusinvariant_prime_divisors(B3)[1])
Toric line bundle on a normal toric variety

julia> w = generic_section(W)
3997*x1^2 - 667*x1*x2 + 2637*x1*x3 - 7466*x1*x4 - 9531*x2^2 + 5011*x2*x3 + 6901*x2*x4 + 5714*x3^2 - 3170*x3*x4 - 4812*x4^2
\end{minted}
Since the coefficients in $w$ are chosen pseudo-randomly, they will vary between executions. Definition~1.1.10 and Proposition~1.1.11 in~\cite{CLS:2011} strongly suggest that the vanishing locus $\mathbb{V}(w) \subseteq \mathbb{P}^3$ is not a toric variety.\footnote{A detailed verification requires examining this locus in each affine chart and applying the referenced proposition. We leave this analysis to the interested reader. Notably, \OSCAR provides tools to compute these affine restrictions using its covered scheme architecture.} Despite this, we can still resolve the model using \FTheoryTools.

To proceed, we construct the modified model following the same steps as before:
\begin{minted}[bgcolor=bg,linenos,breaklines]{julia}
julia> Kbar = anticanonical_bundle(B3)
Toric line bundle on a normal toric variety

julia> a10 = generic_section(Kbar);

julia> a21 = generic_section(Kbar^2*W^(-1));

julia> a32 = generic_section(Kbar^3*W^(-2));

julia> a43 = generic_section(Kbar^4*W^(-3));

julia> a6 = zero(cox_ring(B3));

julia> t2 = global_tate_model(B3, [a10, a21 * w, a32 * w^2, a43 * w^3, a6])
Global Tate model over a concrete base

julia> singular_loci(t2)
2-element Vector{...}:
 (Ideal with 1 generator, (0, 0, 1), "I_1")
 (Ideal with 1 generator, (0, 0, 5), "Split I_5")

julia> singular_loci(t2)[2][1]
Ideal generated by
  3997*x1^2 - 667*x1*x2 + 2637*x1*x3 - 7466*x1*x4 - 9531*x2^2 + 5011*x2*x3 + 6901*x2*x4 + 5714*x3^2 - 3170*x3*x4 - 4812*x4^2
\end{minted}

As in the toric case, the first blowup has center $\mathbb{V}(x, y, w)$ and introduces the exceptional divisor $e_1$:
\begin{minted}[bgcolor=bg,linenos,breaklines]{julia}
julia> t2_1 = blow_up(t, ["x", "y", string(w)]; coordinate_name = "e1")
Partially resolved global Tate model over a concrete base

julia> typeof(ambient_space(t2_1))
CoveredScheme{QQField}
\end{minted}
Let us inspect the ambient space of \verb|t2_1| (we use schematic output below to save space):
\begin{minted}[bgcolor=bg,linenos,breaklines]{julia}
julia> ambient_space(t2_1)
Scheme
  over rational field
with default covering
  described by patches
     1: scheme(-(s1//s0)*x_5_1 + x_4_1, ..., ... - 3170*x_3_1 - 481)
     2: scheme((s0//s1)*x_4_1 - x_5_1, ..., ... - 4812*(s0//s1) - (s2//s1)*x_5_1)
     ...
     20: normal, 5-dimensional toric variety
  in the coordinate(s)
     1: [(s1//s0), (s2//s0), x_1_1, ..., x_5_1]
     2: [(s0//s1), (s2//s1), x_1_1, ..., x_5_1]
     ...
     20: [x_1_12, x_2_12, x_3_12, x_4_12, x_5_12, x_6_12]
\end{minted}
As expected, the ambient space of \verb|t2_1| is a scheme defined over the rational numbers $\mathbb{Q}$.\footnote{Toric geometry typically uses $\mathbb{C}$ as the base ring, but \OSCAR and many other computer algebra systems conventionally use $\mathbb{Q}$ instead. Behind the scenes, \OSCAR converts a toric variety into a scheme while preserving the base ring, leading to a covered scheme over $\mathbb{Q}$.} The output confirms that the scheme consists of multiple affine charts, with their affine coordinates explicitly provided. However, to avoid excessive verbosity, transition maps are omitted from the printout. We will not delve further into the technical details here. The interested reader can find guidance on how to interpret this information in~\cite{Zach2023SchemesOSCAR} and the \OSCAR documentation on algebraic geometry:
\begin{center}
\href{https://docs.oscar-system.org/stable/AlgebraicGeometry/intro/}{https://docs.oscar-system.org/stable/AlgebraicGeometry/intro/}.
\end{center}

In principle, we could now proceed with the second blowup with center $\mathbb{V}(y, e_1)$. However, aside from being cumbersome, this process presents several technical caveats. Most notably, the scheme ambient space of \verb|t2_1| has no global homogeneous coordinates. This means that neither $e_1$ nor $y$ exist as global coordinates, making it meaningless to define a blowup at the vanishing locus of $e_1$ and $y$. Instead, the vanishing locus of $e_1$ must be understood as the exceptional divisor of the blowdown morphism $b$ from the ambient space of \verb|t2_1| to the ambient space of \verb|t2|. More precisely, we should think of $\{e_1 = 0\}$ via the ideal sheaf associated with the exceptional divisor of the blowdown morphism $b$. Likewise, one constructs the ideal sheaf corresponding to $\{y = 0\}$ in the ambient space of \verb|t2| and computes its strict transform along $b$. Only after this step can the two ideal sheaves be combined to instruct \OSCAR's covered-scheme framework to execute a blowup with center being the covered-scheme analogue of $\mathbb{V}(e_1, y)$ in the ambient space of \verb|t2_1|.

This strategy is clearly more cumbersome than executing a sequence of toric blowups. Rather than detailing this process explicitly and repeating it four times, we follow the same approach as before: we teach the model a resolution and let it execute the full resolution sequence automatically:
\begin{minted}[bgcolor=bg,linenos,breaklines]{julia}
julia> add_resolution!(t2, [["x", "y", "w"], ["y", "e1"], ["x", "e4"], ["y", "e2"], ["x", "y"]], ["e1", "e4", "e2", "e3", "s"])

julia> explicit_model_sections(t2)["w"] = w;

julia> t2_res = resolve(t2, 1)
Partially resolved global Tate model over a concrete base
\end{minted}

One might be tempted to inspect the Tate polynomial:
\begin{minted}[bgcolor=bg,linenos,breaklines]{julia}
tate_polynomial(t2_res)
ERROR: ArgumentError: Reconstruction of Tate polynomial is not implemented
in the non-toric case
\end{minted}
However, just as there are no global coordinates $e_1$, $y$, there is also no longer a Tate polynomial for this model. Instead, the originally singular locus $\mathbb{V}(p_t)$ has been converted into an ideal sheaf, and its strict transform has been computed throughout the sequence of blowups. As a result, rather than querying the Tate polynomial, we must now retrieve the Tate ideal sheaf:
\begin{minted}[bgcolor=bg,linenos,breaklines]{julia}
julia> tate_ideal_sheaf(t2_res)
Sheaf of ideals
  on scheme over QQ covered with 48 patches
     1: [(s1//s0), (s1//s0), (s2//s0), x_1_1, x_2_1, x_3_1]   
        scheme(0, -(s1//s0)*(s1//s0)^2*(s2//s0) + 3997*(s1//s0)*x_1_1^2 - ..., 0, 0)
     2: [(s0//s1), (s2//s0), x_1_1, x_2_1, x_3_1, x_5_1]      
        scheme(0, -(s0//s1)*(s2//s0)*x_5_1^2 + 3997*(s0//s1)*x_1_1^2*x_5_1 - ..., 0, 0)
    ...
    47: [x_1_11, x_2_11, x_3_11, x_4_11, x_5_11, x_6_11]      scheme(-x_4_11*x_5_11 + x_6_11^2)
    48: [x_1_12, x_2_12, x_3_12, x_4_12, x_5_12, x_6_12]      scheme(-x_4_12*x_5_12 + x_6_12^2)
with restrictions
   1: Ideal with 3 generators
   2: Ideal with 2 generators
  ...
  47: Ideal with 4 generators
  48: Ideal with 4 generators
\end{minted}
The output is too long to display in full. However, readers interested in inspecting it can do so interactively. Recall that as part of the \OSCAR tutorials, we provide a \texttt{Jupyter} notebook containing all the code snippets presented in this article. This notebook is available at:
\begin{center}
    \href{https://github.com/HereAround/MartinsOscarTutorials.jl/blob/master/FTheoryToolsPaper.ipynb}{https://github.com/HereAround/MartinsOscarTutorials.jl/blob/master/FTheoryToolsPaper.ipynb}.
\end{center}

\section{Literature Models} \label{sec:LiteratureModels}

The \texttt{LiteratureModels} framework in \FTheoryTools streamlines model exploration, modification, and validation, reducing computational overhead and enabling precise exploration of the vast F-theory landscape. This section demonstrates its capabilities, including automated model retrieval, metadata access, model tuning, and singularity resolution.

\paragraph{Tackling Computational Challenges in F-Theory}

A recurring challenge in string phenomenology is the need to revisit earlier constructions as mathematical techniques evolve. While this process often leads to valuable new insights, it can also be cumbersome, sometimes requiring researchers to manually recalculate extensive prior computations. This inefficiency not only slows progress but also creates barriers for new researchers entering the field.

To address these challenges, \FTheoryTools introduces \texttt{LiteratureModels}, a dedicated framework designed to catalog and facilitate the exploration of well-established F-theory constructions in a computationally accessible format. This comes in the form of a database of existing models from the F-theory literature. By automating essential calculations and performing rigorous consistency checks to minimize errors, \texttt{LiteratureModels} can significantly streamline research efforts. It also integrates key results from the literature, such as topological data and known crepant resolutions, to allow researchers to engage with existing models in a more efficient way.

This approach offers several advantages. Researchers can efficiently explore constructions they may not have previously encountered, broadening their understanding of F-theory models. Instead of laboriously reproducing known results, users can retrieve stored data instantly, reducing computational overhead. The framework allows for targeted searches, enabling models to be filtered based on specific criteria, such as gauge group factors. Once a literature model is loaded, it can be further analyzed and modified using the full suite of methods in \FTheoryTools, facilitating deeper exploration.

The database is being actively expanded and already includes models from a number of papers, such as \cite{Krause:2011xj, Morrison:2012ei, Cvetic:2015, Taylor:2015xtz}. Additionally, it encompasses entire families of models discussed in \cite{Lawrie:2012gg, KMOPR15, CHLLT19}. With continued expansion, we aim to make \texttt{LiteratureModels} an indispensable tool for both experts and newcomers, accelerating research and fostering new discoveries in F-theory.

\paragraph{Building a Robust Infrastructure for Model Exploration}
\label{mardi_sec}
Mathematical structures created in \OSCAR are saved using the MaRDI file format~\cite{mrdi-file-format}, which is being developed as part of the Mathematics Research Data Initiative (MaRDI)~\cite{mardi:whitepaper}. The MaRDI file format is platform-independent (being JSON-based) and aims to establish interoperability across a broad range of computer algebra systems, extending well beyond \OSCAR. Notably, it already supports systems such as Nemo/Hecke~\cite{MR3703682}, GAP~\cite{GAP}, polymake~\cite{polymake:2000}, and Singular~\cite{SINGULAR}, with partial extensions to Macaulay2~\cite{M2}, Magma~\cite{MR1484478}, and Sage~\cite{sagemath}, and further integrations anticipated. 

The first \texttt{LiteratureModels} were developed in parallel with the MaRDI file format. As of now, most models in the database are stored as structured JSON files. This format has proven to be both practical and human-readable, making it a viable option for researchers who may wish to contribute additional models. We are currently in the process of transitioning these JSON files to the MaRDI file format, ensuring compatibility with the wider ecosystem of computer algebra systems. Since the MaRDI format is built upon JSON, this transition is straightforward. Some models have already been converted, including the \emph{F-theory geometry with the most flux vacua}~\cite{Taylor:2015xtz}, which we will study extensively in \cref{sec:g4StressTest}.

Beyond its role in interoperability, the MaRDI file format also facilitates efficient data storage via compression through subtree reduction. While these technical details may not be directly relevant for most users, they ensure that large-scale computations---such as those required for models like~\cite{Taylor:2015xtz}---can be stored and retrieved efficiently without the need for repeated computations.

\paragraph{Standardizing Data and Improving Reproducibility}

We envision the MaRDI data format as a valuable tool for enhancing the transparency, reproducibility, and overall impact of F-theory research. By supplementing new constructions with machine-readable files that generate the underlying geometry, both researchers and reviewers can utilize \FTheoryTools to analyze, modify, and verify these models. This practice not only helps standardize data-sharing across the community---fostering collaboration and promoting further advancements in the field---but may also accelerate the review process. Notably, this approach aligns with ongoing MaRDI initiatives that emphasize the importance of peer-reviewing the software components of mathematical research to ensure accessibility, documentation, and reproducibility~\cite{hanselman2025guidelines}. To further enhance the reliability of \FTheoryTools, future efforts could incorporate formal verification techniques from established proof systems such as \texttt{LEAN}~\cite{deMoura2015lean, leanprover}.

\subsection{Fundamentals of Literature Models}

\paragraph{Loading and Inspecting a Model}

We begin by revisiting the F-theory model with an $\asu(5) \oplus \au(1)$ gauge algebra from~\cite{Krause:2011xj}. This example illustrates how to retrieve a model from the database, explore its properties, and perform an automated resolution. As in \cref{sec:ToricBlowup}, we choose the gauge divisor $W$ to be a toric divisor:
\begin{minted}[bgcolor=bg,linenos,breaklines]{julia}
julia> B3 = projective_space(NormalToricVariety, 3)
Normal toric variety

julia> W = torusinvariant_prime_divisors(B3)[1]
Torus-invariant, prime divisor on a normal toric variety

julia> t = literature_model(arxiv_id = "1109.3454", equation = "3.1", base_space = B3, defining_classes = Dict("w" => W))
Global Tate model over a concrete base -- SU(5)xU(1) restricted Tate model based on arXiv paper 1109.3454 Eq. (3.1)
\end{minted}
In this case, the correct model is extracted using metadata, specifically the arXiv ID and equation number from~\cite{Krause:2011xj}. Notably, the model stores and provides access to bibliographic metadata. For brevity, we present a subset of the stored metadata:
\begin{minted}[bgcolor=bg,linenos,breaklines]{julia}
julia> journal_name(t)
"Nucl. Phys. B"

julia> paper_authors(t)
3-element Vector{String}:
 "Sven Krause"
 "Christoph Mayrhofer"
 "Timo Weigand"

julia> arxiv_doi(t)
"10.48550/arXiv.1109.3454"

julia> journal_model_equation_number(t)
"3.1"

julia> journal_model_page(t)
"9"
\end{minted}
In referencing literature models, we primarily link to arXiv preprints, as these are publicly accessible to all researchers. Note that these versions may differ from the corresponding published journal articles. When available, we also support DOI-based references to the peer reviewed versions. In such cases, internal navigation---such as page or equation numbers---must follow the structure of the journal version. While we aim to support both sources where possible, our database is both large and growing, and discrepancies or omissions may exist. To date, we are not aware of any substantial differences between the preprint and published versions for the included models, though more subtle differences (e.g., corrections) remain a possibility. We plan to handle such cases as they arise, potentially with redirections or warnings.

\paragraph{Divisor Classes and Sections of Literature Models}

F-theory models are defined in terms of divisor classes on the base. For instance, the model we have just constructed has a single defining divisor class---the class of the gauge divisor $W$. Because this is the only defining class for this model, it is the only divisor class we need to supply in order to construct the model. The model is then constructed using a random generic section of the line bundle associated with each defining class; in this case, the section is denoted $w$. The defining classes of a model can be accessed as follows:
\begin{minted}[bgcolor=bg,linenos,breaklines]{julia}
julia> defining_classes(t)
Dict{String, ToricDivisorClass} with 1 entry:
  "w" => Divisor class on a normal toric variety
\end{minted}
Note that the name of the section, $w$, is used as the key in this dictionary to minimize the amount of notation associated with the model.

Since \verb|t| is a global Tate model, it inherently includes Tate sections $a_i$, which in this case factor into products of sections $a_{i, j}$ with powers of the section $w$. All sections associated with a model can be accessed via:
\begin{minted}[bgcolor=bg,linenos,breaklines]{julia}
julia> model_sections(t)
9-element Vector{String}:
 "a21"
 "a6"
 "a3"
 "w"
 "a2"
 "a1"
 "a43"
 "a4"
 "a32"
\end{minted}
We can also inspect how the Tate sections $a_i$ are parametrized by the sections $a_{i,j}$ and the section $w$:
\begin{minted}[bgcolor=bg,linenos,breaklines]{julia}
julia> model_section_parametrization(t)
Dict{String, MPolyRingElem} with 4 entries:
  "a6" => 0
  "a3" => w^2*a32
  "a2" => w*a21
  "a4" => w^3*a43
\end{minted}
This naturally leads to the concept of \emph{tunable sections}---a subset of model sections that can be tuned (further specialized) without altering the model's fundamental structure:
\begin{minted}[bgcolor=bg,linenos,breaklines]{julia}
julia> tunable_sections(t)
5-element Vector{String}:
 "a21"
 "w"
 "a1"
 "a43"
 "a32"
\end{minted}
Such a tuning can be carried out using the \verb|tune| function, which we will discuss shortly.

The divisor classes of the various sections in the model can be easily retrieved in two different forms. The divisor classes of every section associated with the model (as \OSCAR divisor classes) can be retrieved:
\begin{minted}[bgcolor=bg,linenos,breaklines]{julia}
julia> classes_of_model_sections(t)
Dict{String, ToricDivisorClass} with 9 entries:
  "a21" => Divisor class on a normal toric variety
  "a6"  => Divisor class on a normal toric variety
  "a3"  => Divisor class on a normal toric variety
  "w"   => Divisor class on a normal toric variety
  "a2"  => Divisor class on a normal toric variety
  "a1"  => Divisor class on a normal toric variety
  "a43" => Divisor class on a normal toric variety
  "a4"  => Divisor class on a normal toric variety
  "a32" => Divisor class on a normal toric variety
\end{minted}
Alternatively, we can retrieve vectors that express the divisor classes of the tunable sections as linear combinations of the anticanonical class of the base and the defining classes:
\begin{minted}[bgcolor=bg,linenos,breaklines]{julia}
julia> classes_of_tunable_sections_in_basis_of_Kbar_and_defining_classes(t)
Dict{String, Vector{Int64}} with 5 entries:
  "a21" => [2, -1]
  "w"   => [0, 1]
  "a1"  => [1, 0]
  "a43" => [4, -3]
  "a32" => [3, -2]
\end{minted}

In some cases, it is also useful to inspect the explicit polynomial expressions for the model sections:
\begin{minted}[bgcolor=bg,linenos,breaklines]{julia}
julia> explicit_model_sections(t)
Dict{String, MPolyDecRingElem{QQFieldElem, QQMPolyRingElem}} with 9 entries:
  "a21" => -2730*x1^7 - 4171*x1^6*x2 + 4757*x1^6*x3 + 6095*x1^6*x4 - ...
  "a6"  => 0
  "a3"  => -9830*x1^12 - 8294*x1^11*x2 + 5284*x1^11*x3 + 1465*x1^11*x4 + ...
  "w"   => x1
  "a2"  => -2730*x1^8 - 4171*x1^7*x2 + 4757*x1^7*x3 + 6095*x1^7*x4 - ...
  "a1"  => -7462*x1^4 - 8047*x1^3*x2 - 5046*x1^3*x3 - 1889*x1^3*x4 - ...
  "a43" => 7748*x1^13 + 9216*x1^12*x2 - 2115*x1^12*x3 + 8057*x1^12*x4 + ...
  "a4"  => 7748*x1^16 + 9216*x1^15*x2 - 2115*x1^15*x3 + 8057*x1^15*x4 + ...
  "a32" => -9830*x1^10 - 8294*x1^9*x2 + 5284*x1^9*x3 + 1465*x1^9*x4 + ...
\end{minted}

\paragraph{Automated Model Resolution}

A key feature of \texttt{LiteratureModels} is the ability to resolve singularities automatically using known blowup sequences. Unlike in \cref{sec:resolutions}, our literature model \verb|t| has a pre-recorded resolution:
\begin{minted}[bgcolor=bg,linenos,breaklines]{julia}
julia> resolutions(t)
1-element Vector{Tuple{Vector{Vector{String}}, Vector{String}}}:
 ([["x", "y", "w"], ["y", "e1"], ["x", "e4"], ["y", "e2"], ["x", "y"]], ["e1", "e4", "e2", "e3", "s"])

julia> t_res = resolve(t, 1)
Partially resolved global Tate model over a concrete base -- SU(5)xU(1) restricted Tate model based on arXiv paper 1109.3454
\end{minted}
This automated resolution minimizes manual workload, especially for complex models. To verify that indeed the correct blowup sequence has been applied, we can inspect the Cox ring of the resolved model:
\begin{minted}[bgcolor=bg,linenos,breaklines]{julia}
julia> v = ambient_space(t_res);

julia> cox_ring(v)
Multivariate polynomial ring in 12 variables over QQ graded by
  x1 -> [1 0 0 0 0 0 0]
  x2 -> [0 1 0 0 0 0 0]
  x3 -> [0 1 0 0 0 0 0]
  x4 -> [0 1 0 0 0 0 0]
  x -> [0 0 1 0 0 0 0]
  y -> [0 0 0 1 0 0 0]
  z -> [0 0 0 0 1 0 0]
  e1 -> [0 0 0 0 0 1 0]
  e4 -> [0 0 0 0 0 0 1]
  e2 -> [-1 -3 -1 1 -1 -1 0]
  e3 -> [0 4 1 -1 1 0 -1]
  s -> [2 6 -1 0 2 1 1]

julia> M = matrix(map_from_torusinvariant_weil_divisor_group_to_class_group(v));

julia> MT = transpose(M)
[1   0   0   0   0   0   0   0   0   -1    0    2]
[0   1   1   1   0   0   0   0   0   -3    4    6]
[0   0   0   0   1   0   0   0   0   -1    1   -1]
[0   0   0   0   0   1   0   0   0    1   -1    0]
[0   0   0   0   0   0   1   0   0   -1    1    2]
[0   0   0   0   0   0   0   1   0   -1    0    1]
[0   0   0   0   0   0   0   0   1    0   -1    1]
\end{minted}
Note that the columns of this matrix give the gradings of the variables in the Cox ring. Applying elementary row transformations to the grading matrix \verb|MT| recovers:
\begin{center}
\begin{tabular}{c c c c | c c c | c c c c c} 
 \hline
 $x_1$ & $x_2$ & $x_3$ & $x_4$ & $x$ & $y$ & $z$ & $e_1$ & $e_4$ & $e_2$ & $e_3$ & $s$ \\\hline
    1 & 1 & 1 & 1 & 8 & 12 & 0 &  0 &  0 &  0 &  0 &  0 \\
    0 & 0 & 0 & 0 & 2 &  3 & 1 &  0 &  0 &  0 &  0 &  0 \\\hline
    1 & 0 & 0 & 0 & 1 &  1 & 0 & -1 &  0 &  0 &  0 &  0 \\
    0 & 0 & 0 & 0 & 0 &  1 & 0 &  1 & -1 &  0 &  0 &  0 \\
    0 & 0 & 0 & 0 & 1 &  0 & 0 &  0 &  1 & -1 &  0 &  0 \\
    0 & 0 & 0 & 0 & 0 &  1 & 0 &  0 &  0 &  1 & -1 &  0 \\
    0 & 0 & 0 & 0 & 1 &  1 & 0 &  0 &  0 &  0 &  0 & -1 \\\hline
\end{tabular}
\end{center}
The last five rows of this matrix show that the desired blowup sequence has indeed been executed.

This process extends naturally to non-toric resolutions, as demonstrated in \cref{sec:NonToricBlowup}. Below, we present the complete code implementation using literature models. Compared to our initial discussion in \cref{sec:NonToricBlowup}, the literature model approach requires significantly less manual input. This highlights the efficiency gains that \texttt{LiteratureModels} provides, making complex model exploration and resolution more accessible and systematic.
\begin{minted}[bgcolor=bg,linenos,breaklines]{julia}
julia> B3 = projective_space(NormalToricVariety, 3)
Normal toric variety

julia> W = 2 * torusinvariant_prime_divisors(B3)[1]
Torus-invariant, prime divisor on a normal toric variety

julia> t = literature_model(arxiv_id = "1109.3454", equation = "3.1", base_space = B3, defining_classes = Dict("w" => W))
Global Tate model over a concrete base -- SU(5)xU(1) restricted Tate model
based on arXiv paper 1109.3454 Eq. (3.1)

julia> t_res = resolve(t, 1)
Partially resolved global Tate model over a concrete base -- SU(5)xU(1)
restricted Tate model based on arXiv paper 1109.3454 Eq. (3.1)
\end{minted}
This example marks the conclusion of our general exposition, setting the stage for more specialized applications of literature models.

\subsection{Toric Hypersurface Fibrations}

\label{subsec:FTheoryOnAllToricHypersurfaces}

For a further set of examples, we can consider the toric hypersurface fibrations. These models, studied extensively in~\cite{Braun:2013nqa, KMOPR15}, associate a smooth toric variety $\mathbb{P}_{F_i}$ to each of the 16 two-dimensional reflexive polyhedra $F_i$, where $i = 1,\ldots,16$. Each toric variety $\mathbb{P}_{F_i}$ admits a corresponding Calabi--Yau hypersurface, which is a genus-one curve $\mathcal{C}_{F_i}$. These curves are obtained as the vanishing locus of a generic section of the anti-canonical bundle $\overline{K}_{\mathbb{P}_{F_i}}$ and can be constructed systematically via the Batyrev construction~\cite{Batyrev:1994hm}. The Calabi--Yau $X_{F_i}$ is then constructed by fibering the curve $\mathcal{C}_{F_i}$ over a concretley chosen (smooth) base $B$. When the toric ambient space of the fiber is fully resolved, the resulting fibration is smooth. Further technical details on the construction can be found in \cite{Batyrev:1994hm, Braun:2013nqa, KMOPR15}.

One of the main reasons toric hypersurface fibrations are of particular interest is their direct realization of phenomenologically relevant gauge groups and matter representations. Notably, they naturally encode the gauge symmetries of the Standard Model ($X_{F_{11}}$), the Pati--Salam model ($X_{F_{13}}$), and the trinification model ($X_{F_{16}}$). Moreover, these models are related naturally via Higgsing chains: for instance, the $X_{F_{11}}$ can be obtained via Higgsing of the $X_{F_{13}}$ or $X_{F_{16}}$ models, as detailed in the appendix of~\cite{KMOPR15}.

To make these models readily accessible for further research, we have implemented all 16 toric hypersurface fibrations described in~\cite{KMOPR15}, along with their corresponding Weierstrass models, as literature models within \FTheoryTools. Information such as their zero sections, Mordell--Weil generating sections, and torsion sections is included in our framework. Let us now construct and analyze an explicit example from this class, demonstrating how \texttt{LiteratureModels} facilitate their exploration.

As an example, let us use \OSCAR to construct the hypersurface model with fiber ambient space $F_{11}$. For the sake of simplicity, we will construct this model over the toric base $\PP^3$. Recall that this model realizes the gauge algebra of the Standard Model. To find information on this model within our database, let us search for models with (at least) this gauge algebra:
\begin{minted}[bgcolor=bg,linenos,breaklines]{julia}
julia> display_all_literature_models(Dict("gauge_algebra" => ["su(3)", "su(2)", "u(1)"]))
Model 33:
Dict{String, Any}("journal_section" => "3", "arxiv_page" => "67", "arxiv_id" => "1408.4808", "gauge_algebra" => Any["su(3)", "su(2)", "u(1)"], "arxiv_version" => "2", "journal_equation" => "3.141", "journal_page" => "67", "arxiv_equation" => "3.142", "journal_doi" => "10.1007/JHEP01(2015)142", "arxiv_section" => "3", "journal" => "JHEP", "file" => "model1408_4808-11-WSF.json", "arxiv_doi" => "10.48550/arXiv.1408.4808", "model_index" => "33", "type" => "weierstrass")

Model 34:
Dict{String, Any}("journal_section" => "3", "arxiv_page" => "67", "arxiv_id" => "1408.4808", "gauge_algebra" => Any["su(3)", "su(2)", "u(1)"], "arxiv_version" => "2", "journal_equation" => "3.141", "journal_page" => "67", "arxiv_equation" => "3.142", "journal_doi" => "10.1007/JHEP01(2015)142", "arxiv_section" => "3", "journal" => "JHEP", "file" => "model1408_4808-11.json", "arxiv_doi" => "10.48550/arXiv.1408.4808", "model_index" => "34", "type" => "hypersurface")

Model 39:
Dict{String, Any}("journal_section" => "3", "arxiv_page" => "75", "arxiv_id" => "1408.4808", "gauge_algebra" => Any["su(3)", "su(2)", "su(2)", "u(1)"], "arxiv_version" => "2", "journal_equation" => "3.167", "journal_page" => "75", "arxiv_equation" => "3.168", "journal_doi" => "10.1007/JHEP01(2015)142", "arxiv_section" => "3", "journal" => "JHEP", "file" => "model1408_4808-14-WSF.json", "arxiv_doi" => "10.48550/arXiv.1408.4808", "model_index" => "39", "type" => "weierstrass")

Model 40:
Dict{String, Any}("journal_section" => "3", "arxiv_page" => "75", "arxiv_id" => "1408.4808", "gauge_algebra" => Any["su(3)", "su(2)", "su(2)", "u(1)"], "arxiv_version" => "2", "journal_equation" => "3.167", "journal_page" => "75", "arxiv_equation" => "3.168", "journal_doi" => "10.1007/JHEP01(2015)142", "arxiv_section" => "3", "journal" => "JHEP", "file" => "model1408_4808-14.json", "arxiv_doi" => "10.48550/arXiv.1408.4808", "model_index" => "40", "type" => "hypersurface")

Model 45:
Dict{String, Any}("journal_section" => "", "arxiv_page" => "2", "arxiv_id" => "1903.00009", "gauge_algebra" => Any["su(3)", "su(2)", "u(1)"], "arxiv_version" => "3", "journal_equation" => "2", "journal_page" => "2", "arxiv_equation" => "2", "journal_doi" => "10.1103/PhysRevLett.123.101601", "arxiv_section" => "II", "journal" => "Physical Review Letters", "file" => "model1903_00009.json", "arxiv_doi" => "10.48550/arXiv.1903.00009", "model_index" => "45", "type" => "hypersurface")
\end{minted}
By inspection, models 33 and 34 have the desired gauge group. They differ in the model type---model 34 is the hypersurface model, whereas model 33 is the corresponding Weierstrass model. We focus here on model 34. We can use its meta data information displayed above to construct this model 34 as follows:
\begin{minted}[bgcolor=bg,linenos,breaklines]{julia}
julia> B3 = projective_space(NormalToricVariety, 3)
Normal toric variety

julia> Kbar = anticanonical_divisor_class(B3)
Divisor class on a normal toric variety

julia> foah11_B3 = literature_model(arxiv_id = "1408.4808", equation = "3.142", type = "hypersurface", base_space = B3, defining_classes = Dict("s7" => Kbar, "s9" => Kbar))
Hypersurface model over a concrete base

julia> model_description(foah11_B3)
"F-theory hypersurface model with fiber ambient space F_11"
\end{minted}
Alternatively, we could also directly employ the model index 34:
\begin{minted}[bgcolor=bg,linenos,breaklines]{julia}
julia> foah11_B3 = literature_model(34, base_space = B3, defining_classes = Dict("s7" => Kbar, "s9" => Kbar))
Hypersurface model over a concrete base
\end{minted}
Now we can investigate some properties of this model. Let us begin with the zero section and its generating sections:\footnote{This construction includes pseudo-random choices for coefficients. The output may therefore differ between executions.}
\begin{minted}[bgcolor=bg,linenos,breaklines]{julia}
julia> zero_section(foah11_B3)
7-element Vector{MPolyDecRingElem{QQFieldElem, QQMPolyRingElem}}:
 1
 0
 -4958*x1^4 + 5651*x1^3*x2 - 1858*x1^3*x3 + 1048*x1^3*x4 - ...
 1
 1
 6656*x1^4 - 6983*x1^3*x2 + 801*x1^3*x3 + 9892*x1^3*x4 - ...
 1

julia> generating_sections(foah11_B3)
1-element Vector{Vector{MPolyDecRingElem{QQFieldElem, QQMPolyRingElem}}}:
 [7678*x1^4 - 4541*x1^3*x2 + ..., 1, 1, -1886*x1^4 + 5209*x1^3*x2 + ..., 1, 1, 0]
\end{minted}
The output lists a single generating section, from which it follows that the Mordell--Weil group of $X_{F_{11}}$ has (at least) rank one. It is important to emphasize that this information is not computed by \FTheoryTools. As of this writing, the software does not include functionality to determine generating sections. Instead, the data was extracted from the original paper and hard-coded into our database. Thus, while the output shows one generating section, this reflects prior knowledge rather than an independent computation. 

Likewise, the gauge algebra was extracted from the paper and is accessible through our database. Let us emphasize that the returned object is provided by the Lie algebra functionality available in \OSCAR. As such, many mechanisms exist to further study these algebras, their attributes and properties. We leave further exploration of this area to the interested reader and merely illustrate the extraction of the gauge algebra\footnote{Note that the gauge summands typically associated with a model are the compact real forms of those returned by this function.} of the above F-theory model:
\begin{minted}[bgcolor=bg,linenos,breaklines]{julia}
julia> g = gauge_algebra(foah11_B3)
Direct sum Lie algebra
  of dimension 12
with summands
  sl_3
  sl_2
  linear Lie algebra
over field of algebraic numbers
\end{minted}

\paragraph{Global Gauge Group Structure}

Beyond the gauge algebra, we have also computed and implemented the discrete global structure of the gauge groups associated with these models. These discrete quotients are not always explicitly specified in~\cite{KMOPR15}, but they are detectable already in the matter spectra discussed there. Our implementation encodes these quotient structures as a nested list of matrices, where each list corresponds to a specific gauge summand in the decomposition provided by \verb|gauge_algebra|. The matrices in one such list belong to the center of the universal covering Lie group of the respective gauge algebra summand. The gauge group is then determined by taking a quotient of this universal covering group by the action of these center elements. 

Together, this list of central elements and the output of \verb|gauge_algebra| fully characterize the global gauge group of a given model. As an illustrative example, we provide the global gauge quotient structure of the toric hypersurface model with fiber ambient space $F_{11}$, whose gauge group is given by
\begin{equation}
    \frac{\SU(3) \times \SU(2) \times \U(1)}{\mathbb{Z}_6}\,.
\end{equation}
Here is how this information is currently represented within \texttt{LiteratureModels}:
\begin{minted}[bgcolor=bg,linenos,breaklines]{julia}
julia> global_gauge_group_quotient(foah11_B3)
3-element Vector{Vector{String}}:
 ["diagonal_matrix(root_of_unity(C,3),3)"]
 ["-identity_matrix(C,2)"]
 ["diagonal_matrix(root_of_unity(C,6,-1),1)"]
\end{minted}
Note that \verb|global_gauge_group_quotient| returns strings rather than matrix objects. This is because, in the current implementation, these matrices are intended to be purely descriptive. Once \OSCAR includes support for Lie groups over infinite fields, we plan to provide dedicated functionality for handling such quotients. However, if one wishes to manipulate these matrices in the current setting, this can be done using the following code:
\begin{minted}[bgcolor=bg,linenos,breaklines]{julia}
julia> C = algebraic_closure(QQ)
Algebraic closure of rational field

julia> diagonal_matrix(root_of_unity(C,3),3)
[Root -0.50+0.87*im of x^2+x+1   Root 0 of x                     Root 0 of x]
[Root 0 of x                     Root -0.50+0.87*im of x^2+x+1   Root 0 of x]
[Root 0 of x                     Root 0 of x                     Root -0.50+0.87*im of x^2+x+1]
\end{minted}
In future developments of \FTheoryTools, we aim to further refine this functionality by representing the gauge group as a single unified object, as matrix groups over the complex numbers have not been implemented in \OSCAR yet.

\paragraph{Relating Hypersurface and Weierstrass Models}

As mentioned before, the Weierstrass model associated with the model \verb|foah11_B3| that we have constructed above is also known to our database. Based on \verb|foah11_B3|, we can extract this Weierstrass model easily with the following command:
\begin{minted}[bgcolor=bg,linenos,breaklines]{julia}
julia> foah11_B3_weier = weierstrass_model(foah11_B3);
\end{minted}
This is part of a larger effort within \FTheoryTools, to remember relations among models. In the case at hand, \verb|foah11_B3_weier| and \verb|foah11_B3| are birationally equivalent models. While this area is a work in progress, this information is indeed known to \FTheoryTools:
\begin{minted}[bgcolor=bg,linenos,breaklines]{julia}
julia> birational_literature_models(foah11_B3)
1-element Vector{String}:
 "1408_4808-11-WSF"
\end{minted}

\paragraph{Model Tuning}

We conclude this exposition with a demonstration of the \verb|tune| method mentioned above. Higgsing and unHiggsing are important physical transitions between models; in the F-theory context, unHiggsings are typically simpler, as they correspond to tunings (or specializations), whereas Higgsings correspond to deformations (or generalizations). Let us see how we can use the \verb|tune| function to explicitly see the unHiggsing $X_{F_{11}} \to X_{F_{13}}$ discussed in~\cite{KMOPR15}, starting with the Weierstrass model of $X_{F_{11}}$ that we constructed above:
\begin{minted}[bgcolor=bg,linenos,breaklines]{julia}
julia> singular_loci(foah11_B3_weier)
3-element Vector{...}:
 (Ideal with 1 generator, (0, 0, 1), "I_1")
 (Ideal with 1 generator, (0, 0, 2), "Non-split I_2")
 (Ideal with 1 generator, (0, 0, 3), "Split I_3")

julia> tunable_sections(foah11_B3_weier)
6-element Vector{String}:
 "s1"
 "s5"
 "s6"
 "s2"
 "s9"
 "s3"

julia> R = cox_ring(base_space(foah11_B3_weier));
 
julia> tuned_model = tune(foah11_B3_weier, Dict("s5"=>zero(R)))
Weierstrass model over a concrete base

julia> singular_loci(tuned_model)
4-element Vector{...}:
 (Ideal with 1 generator, (0, 0, 1), "I_1")
 (Ideal with 1 generator, (0, 0, 2), "Non-split I_2")
 (Ideal with 1 generator, (0, 0, 2), "Non-split I_2")
 (Ideal with 1 generator, (0, 0, 4), "Split I_4")
\end{minted}
Here, we see that the gauge algebra of the $X_{F_{11}}$ model is $\asu(3) \oplus \asu(2) \oplus \au(1)$, as expected.\footnote{You may note that the $I_2$ singularity associated with the $\asu(2)$ summand is non-split; this is not an issue, as $\asu(2) \cong \asp(1)$.} We then list the tunable sections of the model; in this case, we want to tune $s_5 \to 0$, which we carry out using the \verb|tune| command. Using the command \verb|singular_loci| on the newly tuned model then verifies that it has the Pati--Salam gauge algebra, $\asu(4) \oplus \asu(2) \oplus \asu(2)$, as expected of the $X_{F_{13}}$ model.

\section{Enumeration of \texorpdfstring{$G_4$}{G4}-Fluxes}
\label{sec:EnumerationOfG4Fluxes}

In this section, we discuss the enumeration of $G_4$-fluxes within a given F-theory model. This task is fundamental in F-theory model building, and as a result, extensive toolkits have been developed to facilitate such computations~\cite{oai:arXiv.org:1111.1232, Krause:2012yh, Braun:2013nqa, Cvetic:2013uta, Cvetic:2015txa, Lin:2015qsa, Lin:2016vus, Jefferson:2022xft}. As a rigorous test of \FTheoryTools, we will employ both established and novel computational techniques to analyze one of the most complex known F-theory setups, first identified in~\cite{Candelas:1997eh,Lynker:1998pb} and discussed extensively in~\cite{Taylor:2015xtz}.

Before delving into this challenging example, we first demonstrate the capabilities of \FTheoryTools in a simpler geometric setting introduced in \cite{CHLLT19}. This allows for a direct comparison between theoretical expectations and computational results, providing a benchmark for our methodology.

Throughout this section, we assume that $\widehat{Y}_4$ is a smooth, Calabi--Yau fourfold, obtained as a crepant resolution of a singular elliptic fibration $Y_4$, and realized as a hypersurface \begin{equation}
    \widehat{Y}_4 = \mathbb{V}(p) \subseteq X_\Sigma
\end{equation}
within a complete, smooth toric variety $X_\Sigma$.\footnote{In principle, we could only require that the hypersurface be smooth and the ambient space be simplicial; however, we will be computing the characteristic classes of the ambient space and restricting these to the hypersurface, and so we further require that the ambient space be fully resolved to facilitate these calculations.} Note that we are considering elliptic fibrations, rather than genus-one or torus fibrations, so we assume the existence of a (zero) section of $\widehat{Y}_4$.


\subsection{Computation of \texorpdfstring{$G_4$}{G4}-Fluxes} \label{subsec:ComputingG4s}

\subsubsection{A Generating Set for Ambient Vertical \texorpdfstring{$G_4$}{G4}-Fluxes} \label{subsubsec:GensOfG4s}

A $G_4$-flux is an element of $H^{2, 2}(\widehat{Y}_4, \mathbb{R}) \equiv H^{2, 2}(\widehat{Y}_4, \mathbb{C}) \cap H^4(\widehat{Y}_4, \mathbb{R})$ subject to the quantization condition
    \begin{equation}
        G_4 + \frac{1}{2} c_2(\widehat{Y}_4) \in H^4(\widehat{Y}_4, \mathbb{Z})
        \label{eq:M2quantization}
    \end{equation}
as well as several additional physical consistency conditions. As a result of the quantization condition, a valid choice of flux will always have rational coefficients. For this reason, in this paper we will always work with elements of $H^{2, 2}(\widehat{Y}_4, \mathbb{Q}) \equiv H^{2, 2}(\widehat{Y}_4, \mathbb{C}) \cap H^4(\widehat{Y}_4, \mathbb{Q})$; this choice is also computationally advantageous. We will refer to an element of $H^{2, 2}(\widehat{Y}_4, \mathbb{Q})$ that has not been confirmed to satisfy the quantization or other consistency conditions as a ``candidate'' $G_4$-flux.\footnote{Ultimately, we will only be able to check a necessary but not sufficient set of conditions for a flux to satisfy \cref{eq:M2quantization}, so we will always be working with flux candidates.}

The cohomology $H^{2, 2}(\widehat{Y}_4, \mathbb{C})$ enjoys an orthogonal decomposition:
    \begin{equation}
        H^{2, 2}(\widehat{Y}_4, \mathbb{C}) = H^{2, 2}_\text{hor}(\widehat{Y}_4, \mathbb{C}) \oplus H^{2, 2}_\text{vert}(\widehat{Y}_4, \mathbb{C}) \oplus H^{2, 2}_\text{rem}(\widehat{Y}_4, \mathbb{C})\,.
    \end{equation}
Here, the horizontal subspace $H^{2, 2}_\text{hor}(\widehat{Y}_4, \mathbb{C})$ comes from the complex structure variation of the unique harmonic $(4, 0)$-form $\Omega$. The vertical subspace $H^{2, 2}_\text{vert}(\widehat{Y}_4, \mathbb{C})$ is spanned by products of harmonic $(1,1)$-forms:
    \begin{equation}
        H^{2, 2}_\text{vert}(\widehat{Y}_4, \mathbb{C}) = \Span\left(H^{1, 1}(\widehat{Y}_4, \mathbb{C}) \wedge H^{1, 1}(\widehat{Y}_4, \mathbb{C})\right)\,.
    \end{equation}
Components belonging to neither the horizontal nor vertical subspaces are referred to as remainder flux. In this paper, we will concern ourselves with the computation of $H^{2, 2}_\text{vert}(\widehat{Y}_4, \mathbb{Q})$.

In many F-theory settings, explicitly determining a basis or generating set for $H^{2, 2}_\text{vert}(\widehat{Y}_4, \mathbb{Q})$ is computationally prohibitive. However, when $\widehat{Y}_4$ is a hypersurface (or complete intersection) in a complete, simplicial toric ambient space $X_\Sigma$---as in the models studied in~\cite{Taylor:2015xtz, CHLLT19}---a subset of $H^{2, 2}_\text{vert}(\widehat{Y}_4, \mathbb{Q})$ is readily accessible. Specifically, we have:
\begin{equation}
    \left.H^{2, 2}(X_\Sigma, \mathbb{Q}) \right|_{\widehat{Y}_4} \subseteq H^{2, 2}_\text{vert}(\widehat{Y}_4, \mathbb{Q})\,.
\end{equation}
We will refer to fluxes in this subspace as ``ambient'' vertical fluxes.

Theorem~9.3.2 of~\cite{CLS:2011} states that for complete, simplicial toric varieties, $H^{p, q}(X_\Sigma, \mathbb{Q}) = 0$ if $p \ne q$. Consequently, $H^{2, 2}(X_\Sigma, \mathbb{Q}) = H^4(X_\Sigma, \mathbb{Q})$. Furthermore, Theorem~12.4.1 of~\cite{CLS:2011} establishes that $H^4(X_\Sigma, \mathbb{Q})$ is isomorphic to $R_{\mathbb{Q}}(\Sigma)_2$, the vector space of all degree-2 monomials (with respect to standard grading) in the cohomology ring $R_{\mathbb{Q}}(\Sigma)$ of $X_\Sigma$, which then provides a basis for expanding elements of $H^{2, 2}(X_\Sigma, \mathbb{Q})$. The ring $R_{\mathbb{Q}}(\Sigma)$ itself is the quotient of the ring $\mathbb{Q}[x_1, \dots, x_r]$, where $r$ is the number of rays in $\Sigma$, by the sum of the ideal of linear relations and the Stanley--Reisner ideal.

To construct a generating set for the candidate ambient vertical $G_4$-fluxes, we thus proceed as follows:
\begin{enumerate}
    \item Identify a basis for the vector space of degree-2 monomials in the cohomology ring $R_{\mathbb{Q}}(\Sigma)$.
    \item Restrict these basis elements to $\widehat{Y}_4$.
\end{enumerate}

Among these steps, the second is the most computationally challenging. Let $g_1, g_2 \in H^{2, 2}(X_\Sigma, \mathbb{Q})$ be two basis elements obtained in the first step. Then it is possible that $\left. g_1 \right|_{\widehat{Y}_4} = 0$, or that $\left. g_1 \right|_{\widehat{Y}_4} = \left. g_2 \right|_{\widehat{Y}_4}$. However, due to the geometric complexity of the hypersurface $\widehat{Y}_4$, determining these relations is often computationally infeasible in full generality. Consequently, our algorithm in \FTheoryTools only checks whether $\left. g_1 \right|_{\widehat{Y}_4}$ is identically zero by treating $g_1$ as an algebraic cycle and intersecting it with $\widehat{Y}_4$. Currently, all other potential dependencies and trivialities are ignored. Details on the employed algorithm can be found in \cref{subsec:AlgoDetailsRestriction}. 

As a result, our approach yields a generating set for $\left. H^{2, 2}(X_\Sigma, \mathbb{Q}) \right|_{\widehat{Y}_4}$ rather than a basis. This minor trade-off allows us to efficiently expand $G_4$-fluxes using the generating set computed by \FTheoryTools.

Having established the theoretical framework, we now illustrate its computational realization using a QSM model~\cite{CHLLT19}. These F-theory models are classified into 708 families. Each family corresponds to one of the 4319 reflexive three-dimensional lattice polytopes $\Delta$, originally classified in~\cite{KS98}. The specific models within each family are uniquely associated with the fine, regular, star triangulations of the associated polytope $\Delta$.

One such triangulation from each family is readily available in the \FTheoryTools database, enabling the straightforward construction of a representative model for each polytope class. For the purpose of this exposition, we focus on the representative model associated with $\Delta_{283}$:
\begin{minted}[bgcolor=bg,linenos,breaklines]{julia}
julia> qsm_model = literature_model(arxiv_id = "1903.00009", model_parameters = Dict("k" => 283))
Hypersurface model over a concrete base

julia> cox_ring(ambient_space(qsm_model))
Multivariate polynomial ring in 14 variables over QQ graded by
  x1 -> [1 0 0 0 0 0 0 0 0]
  x2 -> [0 1 0 0 0 0 0 0 0]
  x3 -> [0 0 1 0 0 0 0 0 0]
  x4 -> [0 0 0 1 0 0 0 0 0]
  x5 -> [0 1 0 -1 0 0 0 0 0]
  x6 -> [1 0 0 0 0 0 0 0 0]
  x7 -> [-2 1 1 1 0 0 0 0 0]
  v -> [0 0 0 0 1 0 0 0 0]
  e3 -> [0 0 0 0 0 1 0 0 0]
  e2 -> [0 0 0 0 0 0 1 0 0]
  u -> [0 0 0 0 0 0 0 1 0]
  e4 -> [0 0 0 0 0 0 0 0 1]
  e1 -> [0 0 0 0 1 1 0 -1 -2]
  w -> [0 0 0 0 0 1 1 1 1]

julia> stanley_reisner_ideal(ambient_space(qsm_model))
Ideal generated by
  x1*x6
  x2*x4
  x2*x5
  x3*x5
  x3*x7
  v*e2
  v*u
  v*e4
  v*e1
  e3*w
  e3*u
  e3*e4
  e3*e1
  x4*x7
  e2*w
  e2*e4
  e2*e1
  u*w
  u*e1
  e4*w
\end{minted}

We begin by computing a basis of $H^{2, 2}(X_\Sigma, \mathbb{Q})$:
\begin{minted}[bgcolor=bg,linenos,breaklines]{julia}
julia> betti_number(ambient_space(qsm_model), 4)
25

julia> cohomology_ring(ambient_space(qsm_model))
Quotient of multivariate polynomial ring in 14 variables x1, x2, x3, x4, ..., w over rational field by ideal with 25 generators
\end{minted}
Here, the cohomology ring has 14 indeterminates, leading to $\binom{14}{2} = 91$ potential degree-2 monomials. However, only 25 pairs contribute, since $H^{2, 2}(X_\Sigma, \mathbb{Q}) \cong \mathbb{Q}^{b_4(X_\Sigma)}$ and $b_4(X_\Sigma) = 25$. The 25 cohomology classes selected by \FTheoryTools are as follows:\footnote{The optional argument \texttt{check} to the function \texttt{basis\textunderscore{}of\textunderscore{}h22\textunderscore{}ambient} determines whether to verify that the toric variety is complete and simplicial. Since the completeness check is computationally expensive and we know by construction that the ambient space is complete, we opt to skip it here.}
\begin{minted}[bgcolor=bg,linenos,breaklines]{julia}
julia> basis_of_h22_ambient(qsm_model, completeness_check = false)
25-element Vector{CohomologyClass}:
 Cohomology class on a normal toric variety given by x6*e4
 Cohomology class on a normal toric variety given by x4*x6
 Cohomology class on a normal toric variety given by x7^2
 Cohomology class on a normal toric variety given by x5*x6
 Cohomology class on a normal toric variety given by x6*x7
 Cohomology class on a normal toric variety given by x7*e2
 Cohomology class on a normal toric variety given by x7*u
 Cohomology class on a normal toric variety given by x7*e4
 Cohomology class on a normal toric variety given by x7*e1
 Cohomology class on a normal toric variety given by x7*w
 Cohomology class on a normal toric variety given by x5*u
 Cohomology class on a normal toric variety given by x4*e4
 Cohomology class on a normal toric variety given by x5*e4
 Cohomology class on a normal toric variety given by e1*w
 Cohomology class on a normal toric variety given by x6*e2
 Cohomology class on a normal toric variety given by x6*u
 Cohomology class on a normal toric variety given by x6*w
 Cohomology class on a normal toric variety given by x4*e2
 Cohomology class on a normal toric variety given by x4*u
 Cohomology class on a normal toric variety given by x4*e1
 Cohomology class on a normal toric variety given by x4*w
 Cohomology class on a normal toric variety given by x5*e2
 Cohomology class on a normal toric variety given by x5*e1
 Cohomology class on a normal toric variety given by x5*w
 Cohomology class on a normal toric variety given by x6*e1
\end{minted}
A generic algorithm computing this basis, which is emplyed by the function \texttt{basis\_of\_h4}, determines $R_{\mathbb{Q}}(\Sigma)_2$ by computing the leading ideal of $R_{\mathbb{Q}}(\Sigma)$ via a Gr{\"o}bner basis computation. However, for complicated toric varieties---such as the geometry studied in~\cite{Taylor:2015xtz}---this approach is impractical due to its computational complexity. We instead employ an alternative algorithm, discussed further in \cref{subsec:AlgoDetailsGenFinding}, that successfully circumvents this bottleneck by avoiding Gr{\"o}bner basis computations entirely.

The final step is the restriction process. The following function computes the cohomology ring, identifies a basis for $H^{2, 2}(X_\Sigma, \mathbb{Q})$, and then performs the elementary restriction tests discussed above (see \cref{subsec:AlgoDetailsRestriction}):
\begin{minted}[bgcolor=bg,linenos,breaklines]{julia}
julia> g4_gens = chosen_g4_flux_gens(qsm_model);

julia> length(g4_gens)
25
\end{minted}
In this case, it so happens that our elementary restriction tests did not eliminate any of the basis elements of $H^{2, 2}(X_\Sigma, \mathbb{Q})$; in \cref{sec:g4StressTest}, we will see an example where a significant number of elements are removed. Let us now examine the first generator in a bit more detail:
\begin{minted}[bgcolor=bg,linenos,breaklines]{julia}
julia> g4_gens[1]
G4-flux candidate
  - Elementary quantization checks: not executed
  - Transversality checks: not executed
  - Non-abelian gauge group: breaking pattern not analyzed
  - Tadpole cancellation check: not computed

julia> cohomology_class(g4_gens[1])
Cohomology class on a normal toric variety given by x6*e4
\end{minted}
We see in the output that none of the quantization and consistency checks have been executed; we now turn to a discussion of these checks.

\subsubsection{Finding Well-Quantized and Transverse \texorpdfstring{$G_4$}{G4}-Fluxes}
\label{subsec:Fluxes-Quantization}

A flux \(G_4 \in H^{2, 2}(\widehat{Y}_4, \mathbb{Q})\) must satisfy the quantization condition~\cite{Witten:1996md}:
\begin{equation}
    G_4 + \frac{1}{2} c_2(\widehat{Y}_4) \in H^{4}(\widehat{Y}_4, \mathbb{Z})\,.
\end{equation}
Finding all solutions to the quantization condition remains a significant challenge. Given \(G_4 \in H^{2, 2}_\text{vert}(X_\Sigma, \mathbb{Q})\)---expressed as a rational combination of the $G_4$-flux candidate generators identified in the previous section---we seek to determine whether its restriction to $\widehat{Y}_4$ satisfies:
\begin{equation}
\left. G_4 \right|_{\widehat{Y}_4} + \frac{1}{2} c_2( \widehat{Y}_4 ) \in \left. H^{2, 2}(X_\Sigma, \mathbb{Q}) \right|_{\widehat{Y}_4} \cap H^4 ( \widehat{Y}_4, \mathbb{Z} )\,.
\end{equation}
Computing \( H^4 ( \widehat{Y}_4, \mathbb{Z} ) \) directly is generally infeasible for most geometries. Consequently, a widely adopted approach, as applied for instance in~\cite{CHLLT19}, involves imposing a series of necessary but not necessarily sufficient conditions. Specifically, let \(H\) be the divisor of the hypersurface \( \widehat{Y}_4 \) in the ambient space and denote by \(D_i\) a basis of all toric divisors of \( X_\Sigma \). Furthermore, let $\hat{c}_2 \in H^{2,2}(X_\Sigma, \mathbb{Q})$ such that $\left. \hat{c}_2 \right|_{\widehat{Y}_4} = c_2( \widehat{Y}_4 )$. A necessary set of conditions for quantization is then
\begin{equation}
    \int_{X_\Sigma} \left(G_4 + \frac{1}{2} \hat{c}_2\right) \wedge \PD{H} \wedge \PD{D_i} \wedge \PD{D_j} \in \mathbb{Z}\,, \quad \forall \; i, j\,,
    \label{equ:IntegralForIntersectionProduct}
\end{equation}
where $\PD{D}$ indicates the Poincar{\'e} dual cohomology class to the divisor $D$. It is these conditions that we check in \FTheoryTools.

An additional set of necessary conditions for quantization conditions is
\begin{equation}
    \int_{X_\Sigma} \left(G_4 + \frac{1}{2} \hat{c}_2\right) \wedge \PD{H} \wedge \PD{S_{\bm{R}}} \in \mathbb{Z}\,, \quad \forall \; \bm{R}\,,
\end{equation}
where $S_{\bm{R}}$ is the matter surface associated with the representation $\bm{R}$. These additional quantization conditions are planned to be implemented in \FTheoryTools in the future.

\paragraph{Intersection Matrix Computation}
We implement an efficient algorithm to compute the intersection numbers between candidate $G_4$-fluxes, toric divisors of the ambient toric fivefold, and the hypersurface \( \widehat{Y}_4 \) inside the toric ambient fivefold. This algorithm is discussed in \cref{subsec:AlgoSophisticatedIntersection}. The Stanley--Reisner ideal of \( X_\Sigma \) simplifies this process: many toric divisors do not intersect in the ambient space. Moreover, we apply a simple yet effective test to determine whether a pair of toric divisors intersects \( \widehat{Y}_4 \) trivially.

For the QSM model associated with the 3-dimensional reflexive polytope \( \Delta_{283} \), these reductions leave us with 85 relevant pairs $D_i, D_j$ of toric ambient space divisors. Thus, we compute $85 \times 25$ intersection numbers among these divisor pairs and our candidate $G_4$ generators, organizing them into the matrix $C \in \matSpace{85}{25}{\mathbb{Z}}$. Likewise, we obtain $85$ intersection numbers among the divisor pairs and $\frac{1}{2} \hat{c}_2$, which we organize into the column vector $\Vec{o} \in \mathbb{Q}^{85}$. We now seek to find all vectors \(\Vec{q} \in \mathbb{Q}^{25}\) such that
\begin{equation}
    \Vec{n} = C \cdot \Vec{q} + \Vec{o}
    \label{equ:WhatWeSeek}
\end{equation}
for some integer vector \( \Vec{n} \in \mathbb{Z}^{85} \).
To solve \cref{equ:WhatWeSeek}, we compute the Smith normal form of \( C \). This involves identifying invertible integer matrices \(T \in \GL(85, \mathbb{Z})\) and \(U \in \GL(25, \mathbb{Z})\), as well as a diagonal matrix \( S \in \matSpace{85}{25}{\mathbb{Z}}\), such that:
\begin{equation}
    S = T \cdot C \cdot U\,.
\end{equation}
Since \( T \) is invertible over \( \mathbb{Z} \), \cref{equ:WhatWeSeek} is equivalent to:
\begin{equation}
    T \cdot \Vec{n} = S \cdot U^{-1} \cdot \Vec{q} + T \cdot \Vec{o} \,.
\end{equation}
Introducing \( \Vec{n}^\prime = T \cdot \Vec{n} \), $\Vec{q}^\prime = U^{-1} \cdot \Vec{q}$, and \( \Vec{o}^\prime = T \cdot \Vec{o} \) we obtain:
\begin{equation}
    \Vec{n}^\prime = S \cdot \Vec{q}^\prime + \Vec{o}^\prime\,.
\end{equation}
Since \( S \) is diagonal, it is simple to read off the solutions. If \( S_{i,i} = 0 \), then \( \Vec{q}^\prime_i \) is unconstrained by our necessary (but not sufficient) quantization checks, meaning \( \Vec{q}^\prime_i \) can take any rational value.\footnote{If \( S_{i,i} = 0 \) and \( \Vec{o}^\prime_i \notin \mathbb{Z} \), no rational solution exists. However, this situation cannot occur, because $\Vec{o}'$ is in the image of $S$.} When \( S_{i,i} \ne 0 \), \( \Vec{q}^\prime_i = \frac{n_i^\prime - o_i^\prime}{S_{ii}} \) for some integer $n_i^\prime$. If \( \Vec{o}^\prime_i \in \mathbb{Z} \), this offset has no effect on the integrality of solutions and can be ignored (operationally, we set integer entries of $\Vec{o}^\prime$ to zero). Finally, the original solution is recovered via \( \Vec{q} = U \cdot \Vec{q}^\prime \).

In practice, the elementary quantization checks are carried out last, \emph{after} reducing the list of candidate $G_4$ generators using additional consistency conditions (to be discussed shortly) and any other constraints the user applies to the set of fluxes. This is because we want to avoid carrying the offset vector $\Vec{o}$ around with us as we compute other constraints.

\paragraph{Additional Consistency Constraints}

In addition to the quantization condition \labelcref{eq:M2quantization}, there are several other consistency conditions that a $G_4$-flux must satisfy. Preserving the minimal amount of supersymmetry in 4D requires that $G_4 \in H^{2, 2}(\widehat{Y}_4, \mathbb{R}) \subset H^4(\widehat{Y}_4, \mathbb{R})$ (which we have already assumed in our previous discussion) and imposes a primitivity condition $G_4 \wedge J = 0$, with $J$ the K{\"a}hler form of $\widehat{Y}_4$. Furthermore, requiring that the $G_4$-flux dualize to a valid 4-form flux in the F-theory vacuum that does not break Poincar{\'e} symmetry imposes the conditions
    \begin{equation}
        \begin{aligned}
            \int_{X_\Sigma}{G_4 \wedge \PD{H} \wedge \PD{\hat{\rho}^*(D^B_i)} \wedge \PD{\hat{\rho}^*(D^B_j)}} &= 0\,, \\
            \int_{X_\Sigma}{G_4 \wedge \PD{H} \wedge \PD{Z} \wedge \PD{\hat{\rho}^*(D^B_j)}} &= 0\,.
        \end{aligned}
        \label{eq:tranversalityConditions}
    \end{equation}
Here, $Z$ is the divisor of the zero section of the fibration, $D^B_i$ denotes a basis of divisors of the base $B$, and $\hat{\rho}$ is the projection map of the resolved ambient space fibration. We refer to these constraints as transversality conditions. These constraints can be checked using a similar approach to that used above for the quantization condition (without need for an offset vector).

For the vacuum to preserve supersymmetry, there is additionally a D3-brane tadpole cancellation condition, which states that the number of spacetime-filling D3-branes must be a non-negative integer:
    \begin{equation}
        N_\text{D3} = \frac{1}{24} \chi(\widehat{Y}_4) - \frac{1}{2} \int_{\hat{Y}_4} G_4 \wedge G_4 \in Z_{\ge 0}\,.
    \end{equation}
The integrality of this quantity is typically assumed to follow from the quantization condition \labelcref{eq:M2quantization}; however, as we only directly check necessary conditions for the $G_4$ flux to satisfy the quantization condition, checking the integrality of $N_\text{D3}$ is also useful to us.

\paragraph{Implementation in \FTheoryTools} These operations are automated in \FTheoryTools. We observe that the computational complexity varies significantly with the geometry of the model. In particular, for the model in~\cite{Taylor:2015xtz}, execution times are considerably longer than for the QSM model at hand. The following computations complete within seconds:
\begin{minted}[bgcolor=bg,linenos,breaklines]{julia}
julia> fg = special_flux_family(qsm_model, completeness_check = false)
Family of G4 fluxes:
  - Elementary quantization checks: satisfied
  - Transversality checks: satisfied
  - Non-abelian gauge group: breaking pattern not analyzed

julia> mat_int = matrix_integral(fg);

julia> size(mat_int)
(25, 17)

julia> mat_rat = matrix_rational(fg);

julia> size(mat_rat)
(25, 0)
\end{minted}
Here, we have used the function \verb|special_flux_family| to find the family of fluxes that passes the elementary quantization and transversality checks; we will soon see how additional arguments to this function can demand additional properties of the returned flux family. The columns of \verb|mat_int| correspond to those $G_4$-flux candidate generators for which only integral multiples are well-quantized. Here, we find 17 such rational combinations. By contrast, \verb|mat_rat| encodes generators that are unconstrained by our quantization checks, thus allowing arbitrary rational multiples, which in this case is empty.

In certain contexts, it is useful to generate a $G_4$-flux candidate that satisfies specific constraints while otherwise being randomly chosen. \FTheoryTools provides two functions for this purpose:
\begin{minted}[bgcolor=bg,linenos,breaklines]{julia}
julia> g4 = random_flux_instance(fg)
G4-flux candidate
  - Elementary quantization checks: satisfied
  - Transversality checks: satisfied
  - Non-abelian gauge group: breaking pattern not analyzed
  - Tadpole cancellation check: not computed

julia> g4_2 = random_flux(qsm_model, completeness_check = false)
G4-flux candidate
  - Elementary quantization checks: satisfied
  - Transversality checks: satisfied
  - Non-abelian gauge group: breaking pattern not analyzed
  - Tadpole cancellation check: not computed
\end{minted}
The former approach, using \verb|random_flux_instance|, returns a random flux from a given family of fluxes, in this case the family \verb|fg| we computed previously. The latter approach, using \verb|random_flux|, returns a random flux satisfying any constraints specified by the optional arguments, combining the steps performed by \verb|special_flux_family| and \verb|random_flux_instance|.

Given such a flux or a flux family as above, one may wish to investigate additional properties. As indicated in the outputs above, key aspects of interest include the breaking of the non-abelian gauge symmetry and the D3-tadpole constraint. We now proceed to discuss these conditions in detail.

\subsubsection{Finding \texorpdfstring{$G_4$}{G4}-Fluxes with Additional Properties} \label{subsec:Fluxes-FurtherProperties}

We can easily extend these computations to restrict to fluxes that do not break the non-abelian gauge group. The relevant constraint on the fluxes is 
    \begin{equation}
        \int_{X_\Sigma}{G_4 \wedge \PD{H} \wedge \PD{\hat{\rho}^*(D^B_i)} \wedge \PD{E_j}} = 0\,,
    \end{equation}
where the $E_i$ are the exceptional divisors introduced by the resolution, corresponding physically to the Cartan generators of the gauge algebra. This is supported within \FTheoryTools:
\begin{minted}[bgcolor=bg,linenos,breaklines]{julia}
julia> fg_not_breaking = special_flux_family(qsm_model, not_breaking = true, completeness_check = false)
Family of G4 fluxes:
  - Elementary quantization checks: satisfied
  - Transversality checks: satisfied
  - Non-abelian gauge group: unbroken

julia> size(matrix_integral(fg_not_breaking))
(25, 1)
\end{minted}
We see that there is only a single generator for all vertical $G_4$-flux candidates that satisfy the elementary quantization and tranversality checks and do not break the non-abelian gauge group. It is worthwhile to inspect this generator in more detail:
\begin{minted}[bgcolor=bg,linenos,breaklines]{julia}
julia> b22 = chosen_g4_flux_gens(qsm_model);

julia> mat_int = matrix_integral(fg_not_breaking);

julia> g4_sample = sum(mat_int[k] * b22[k] for k in 1:length(b22));

julia> cohomology_class(g4_sample)
Cohomology class on a normal toric variety given by -1//5*x5*x6 + 1//30*x5*e2 + 1//15*x5*u + 1//10*x5*e4 + 1//15*x5*e1 + 1//30*x5*w - 11//15*x6*x7 + 2//15*x6*e2 + 4//15*x6*u + 2//5*x6*e4 + 4//15*x6*e1 + 2//15*x6*w - 7//30*x7^2 + 1//15*x7*e2 + 2//15*x7*u + 1//5*x7*e4 + 2//15*x7*e1 + 1//15*x7*w - 1//6*e1*w
\end{minted}
So, any integral multiple of this cohomology class defines a vertical $G_4$-flux candidate that passes the elementary quantization and transversality checks and does not break the non-abelian gauge group. We should compare this flux to the one chosen in~\cite{CHLLT19} (this particular expression for the flux is taken from~\cite{Bies:2021nje}):
\begin{equation}
    G_4 = \frac{-3}{\KB^3} \bigg( 5 [ e_1 ] \wedge [ e_4 ] + \left. \widehat{\pi}^\ast \left( \KB \right) \wedge \left( - 3 [ e_1 ] - 2 [ e_2 ] - 6 [ e_4 ] + \widehat{\pi}^\ast \left( \KB \right) - 4 \left[ u \right] + \left[ v \right] \right) \bigg)  \right|_{\widehat{Y}_4} \,.
\end{equation}
To compare directly, we create this flux by hand:
\begin{minted}[bgcolor=bg,linenos,breaklines]{julia}
julia> kbar3(qsm_model)
30

julia> divs = torusinvariant_prime_divisors(ambient_space(qsm_model));

julia> e1 = cohomology_class(divs[13]);

julia> e2 = cohomology_class(divs[10]);

julia> e4 = cohomology_class(divs[12]);

julia> u = cohomology_class(divs[11]);

julia> v = cohomology_class(divs[8]);

julia> pb_Kbar = cohomology_class(sum(divs[1:7]));

julia> g4_class = (-3) // kbar3(qsm_model) * (5 * e1 * e4 + pb_Kbar * (-3 * e1 - 2 * e2 - 6 * e4 + pb_Kbar - 4 * u + v));

julia> g4_sample2 = g4_flux(qsm_model, g4_class, completeness_check = false);
\end{minted}
For QSM models, these steps are conveniently collected in the function \verb|qsm_flux|:
\begin{minted}[bgcolor=bg,linenos,breaklines]{julia}
julia> qsm_flux(qsm_model)
G4-flux candidate
  - Elementary quantization checks: satisfied
  - Transversality checks: satisfied
  - Non-abelian gauge group: unbroken
  - Tadpole cancellation check: not computed

julia> qsm_flux(qsm_model) == g4_sample2
true
\end{minted}
Let us examine this flux---or, more precisely, its defining cohomology class---in more detail:
\begin{minted}[bgcolor=bg,linenos,breaklines]{julia}
julia> cohomology_class(g4_sample2)
Cohomology class on a normal toric variety given by -3//5*x5*x6 + 1//10*x5*e2 + 1//5*x5*u + 3//10*x5*e4 + 1//5*x5*e1 + 1//10*x5*w - 11//5*x6*x7 + 2//5*x6*e2 + 4//5*x6*u + 6//5*x6*e4 + 4//5*x6*e1 + 2//5*x6*w - 7//10*x7^2 + 1//5*x7*e2 + 2//5*x7*u + 3//5*x7*e4 + 2//5*x7*e1 + 1//5*x7*w - 1//2*e1*w

julia> 3 * cohomology_class(g4_sample) - cohomology_class(g4_sample2)
Cohomology class on a normal toric variety given by 0
\end{minted}
So we see that the $G_4$-flux chosen in~\cite{CHLLT19} is three times the \emph{integral} generator for the flux family computed by \FTheoryTools. We will see the cause for the discrepancy momentarily.

We can of course also compute the number of D3-branes to check the tadpole condition, both for individual $G_4$-fluxes as well as for an entire family of fluxes. Let us focus on \verb|fg_not_breaking|. It has a single integral generator $g$, so the most general such flux candidate is of the form $a_1 \cdot g$, $a_1 \in \mathbb{Z}$. The D3-tadpole for this family of fluxes is then computed with \FTheoryTools as follows:
\begin{minted}[bgcolor=bg,linenos,breaklines]{julia}
julia> d3_tadpole_constraint(fg_not_breaking)
-1//12*a1^2 + 123//4
\end{minted}
We see that this quantity is integral for $a_1 = \pm 3 \mod 12$ and nonnegative for $\abs{a_1} \le 19$. This gives us an additional integrality constraint on the fluxes, beyond those determined by the elementary quantization checks, bringing our result into accord with that in~\cite{CHLLT19}.

As a consistency check, let us directly compute the D3-tadpole for the flux $3 \cdot g$, i.e., make the choice $a_1 = 3$, corresponding to the choice made in~\cite{CHLLT19}. We can do this using the \verb|flux_instance| function, providing a vector of integer coefficients for the integral generators and a vector of rational coefficients (in our case, the empty vector) for the rational generators of our flux family:
  
\begin{minted}[bgcolor=bg,linenos,breaklines]{julia}
julia> g4_exp = flux_instance(fg_not_breaking, [3], [])
G4-flux candidate
  - Elementary quantization checks: satisfied
  - Verticality checks: satisfied
  - Non-abelian gauge group: unbroken
  - Tadpole constraint: not computed

julia> d3_tadpole_constraint(g4_exp)
30

julia> d3_tadpole_constraint(g4_exp) == -1//12*3^2 + 123//4
true
\end{minted}
We see that the result is a positive integer, as desired.

\subsection{\texorpdfstring{$G_4$}{G4}-Fluxes in a Complex Model---A Computational Stress Test}
\label{sec:g4StressTest}

In this section, we extend the computations from the previous section to a significantly more intricate geometry: ``The F-theory geometry with most flux vacua'', first identified in~\cite{Candelas:1997eh,Lynker:1998pb} and described extensively in~\cite{Taylor:2015xtz}. This analysis serves multiple purposes.

First, from a conceptual standpoint, studying $G_4$-fluxes in the model of~\cite{Taylor:2015xtz} through a statistical lens is of great interest. Specifically, given a random $G_4$-flux, to what extent does it break the non-abelian gauge group? We do expect that a generic $G_4$-flux will break some of the gauge factors, but a complete study of this process remains absent in the literature. Our work lays the foundation for addressing this question and provides insights applicable to similarly complicated geometries.

Second, the computational challenge presented by this model is substantial. In the previous section, we determined $G_4$-flux generators and computed all relevant topological intersection numbers efficiently using Gr{\"o}bner basis computations. However, it is well known that the complexity of Gr{\"o}bner basis computations scales double-exponentially with the number of variables. This renders the direct application of such methods to the geometry of~\cite{Taylor:2015xtz} practically infeasible. To overcome this limitation, we propose alternative computational algorithms, pushing the boundaries of what is technically achievable. Consequently, we treat the computations in~\cite{Taylor:2015xtz} as a rigorous stress test for \FTheoryTools.

For the reader's convenience, and to underscore the computational challenges involved, we begin by reviewing the geometry constructed in~\cite{Taylor:2015xtz} and its resolution via a sequence of toric blowups. We then proceed to repeat the computations from the previous section within this complex geometric setting. The most intricate aspect, namely the presentation of algorithms for determining candidate $G_4$-flux generators and computing intersection numbers, is discussed in \cref{sec:AppendixAlgorithms}.

\subsubsection{Review: ``The F-theory geometry with most flux vacua''}
\label{subsubsec:RecapOfBigModel}

The work~\cite{Taylor:2015xtz} identifies an elliptically fibered fourfold that they suggest gives rise to $O(10^{272,000})$ F-theory flux vacua, constituting the overwhelming majority of all F-theory flux vacua. This geometry features a geometrically non-Higgsable gauge group $E_8^9 \times F_4^8 \times (G_2 \times \mathrm{SU}(2))^{16}$ and is described as a global Tate model over a smooth, complete three-dimensional toric variety without a torus factor. The base is defined by a polyhedral fan with 101 rays and 198 maximal cones. Consequently, in \texttt{OSCAR}, the toric variety is covered by 198 affine charts corresponding to these maximal cones. The Cox ring of the base space consists of 101 homogeneous coordinates, labeled as $w_0, w_1, \dotsc, w_{100}$, following the notation of~\cite{Taylor:2015xtz}.

Executing all computations necessary to construct this model from scratch in \FTheoryTools may easily require half an hour on a typical personal computer. The primary computational bottleneck is the determination of the Tate polynomial, which contains 355,785 monomials, rendering its computation exceedingly time-consuming. While future performance improvements in \FTheoryTools are anticipated, this problem remains inherently challenging. We mitigate this computational burden using serialization techniques. Specifically, we provide a precomputed model file, \texttt{1511-03209.mrdi}, available at~\cite{bmtMaRDIFiles}. The following command downloads and utilizes this serialized file automatically, reducing model instantiation time to at most a few minutes on most personal computers:
\begin{minted}[bgcolor=bg,linenos,breaklines]{julia}
julia> t = literature_model(arxiv_id = "1511.03209")
Global Tate model over a concrete base -- The F-theory geometry with most flux vacua based on arXiv paper 1511.03209 Eq. (2.11)
\end{minted}

A resolution sequence for this model can be derived from~\cite{Lawrie:2012gg, Esole:2018mqb, Esole:2017rgz}. We employ a sequence of 206 toric blowups. For instance, over the locus $\mathbb{V}(w_{99}) \subset B_3$, an $E_8$ singularity is resolved by introducing exceptional divisors \texttt{e99\_1}, \texttt{e99\_2}, \textellipsis, \texttt{e99\_14}, as detailed in~\cite{Lawrie:2012gg}. Since all singularities arise over loci of the form $\{ w_i = 0 \}$, we adopt this convention for labeling the exceptional divisors. Afterwards, it proves advantageous for our computational framework to also resolve the remaining singularities of the toric ambient space, which requires three additional toric blowups.

Despite the relative simplicity of toric blowups, the computation of the strict transform of the Tate polynomial---consisting in the case at hand of 355,785 monomials---is tedious and time consuming. Consequently, executing this resolution from scratch may require several hours on a personal computer. Once again, serialization offers a remedy. The following command utilizes a precomputed \texttt{.mrdi} file available at~\cite{bmtMaRDIFiles}, reducing execution time of the resolution to a few minutes on most personal computers:
\begin{minted}[bgcolor=bg,linenos,breaklines]{julia}
julia> t_res = resolve(t, 1)
Partially resolved global Tate model over a concrete base -- The F-theory geometry with most flux vacua based on arXiv paper 1511.03209 Eq. (2.11)
\end{minted}
The second argument, \texttt{1}, specifies which resolution sequence is executed or loaded by the above command. Currently, \FTheoryTools supports only a single blow-up sequence for resolving the model of~\cite{Taylor:2015xtz}; thus, the first resolution sequence is the only applicable choice in the present case.

\subsubsection{Generators of \texorpdfstring{$G_4$}{G4}-Fluxes}

The ambient space of the resolved fourfold $\widehat{Y}_4$, computed in the previous section, is a complete, smooth toric variety defined by a polyhedral fan with 313 rays and 2,836 maximal cones. The hypersurface $\widehat{Y}_4$ itself is given by a polynomial with 355,785 monomials. As in the previous case, we employ the restriction
\begin{equation}
    \left. H^{2, 2}(X_\Sigma, \mathbb{Q}) \right|_{\widehat{Y}_4} \subseteq H^{2, 2}_\text{vert}(\widehat{Y}_4, \mathbb{Q}).
\end{equation}
Thus, our first goal is to compute a basis for $H^{2, 2}(X_\Sigma, \mathbb{Q}) = H^4(X_\Sigma, \mathbb{Q})$ (cf.\ Theorem~9.3.2 of~\cite{CLS:2011}) and subsequently restrict it to the hypersurface $\widehat{Y}_4$.

\OSCAR provides a general-purpose algorithm, \verb|basis_of_h4|, that determines a basis for $H^4(X_\Sigma, \mathbb{Q})$ using Gr{\"o}bner basis computations on the cohomology ring of $X_\Sigma$. However, in the present case, this algorithm is unlikely to terminate within a practical time frame, as we can see by computing the cohomology ring of the ambient space:\footnote{As discussed earlier, \OSCAR supports calculation of the cohomology ring only for complete and simplicial toric varieties (Theorem~12.4.1 in~\cite{CLS:2011}). By default, the method \texttt{cohomology\_ring} verifies these conditions, which introduces a significant computational bottleneck for the ambient space $X_\Sigma$ of~\cite{Taylor:2015xtz}. To circumvent this issue, we disable this check using the optional argument \texttt{completeness$\_$check = false}. The same considerations apply to the Chow ring (Theorem~12.5.3 in~\cite{CLS:2011}), which can, however, be extended to non-complete varieties as the \emph{combinatorial Chow ring}~\cite{Peg14}. This distinction also explains why only the cohomology ring constructor in \texttt{OSCAR} supports the \texttt{completeness$\_$check} argument.}
\begin{minted}[bgcolor=bg,linenos,breaklines]{julia}
julia> amb = ambient_space(t_res);

julia> cohomology_ring(amb, completeness_check = false);
\end{minted}
We suppress the lengthy output from this command here. The cohomology ring is a quotient of a multivariate polynomial ring with 313 indeterminates by an ideal generated by 46,547 relations. Given the sheer complexity of the Gr{\"o}bner basis computation required, waiting for termination of \verb|basis_of_h4| is impractical. \FTheoryTools instead provides a specialized algorithm that significantly outperforms the generic approach. Details can be found in \cref{subsec:AlgoDetailsGenFinding}. In the case of the model at hand, the basis of $H^{2, 2}(X_\Sigma, \mathbb{Q}) = H^4(X_\Sigma, \mathbb{Q})$ consists of 1,109 elements, as is readily verified by a Betti number computation:
\begin{minted}[bgcolor=bg,linenos,breaklines]{julia}
julia> betti_number(amb, 4)
1109
\end{minted}
We restrict the basis elements to $\widehat{Y}_4$ with the algorithm detailed in \cref{subsec:AlgoDetailsRestriction}. For computational efficiency, we execute only computationally simple checks to identify and remove basis elements that are trivial upon restriction to the hypersurface. Computationally more expensive tests to tell whether the restriction of a cohomology class to the hypersurface is trivial or to find relations among restricted basis elements are omitted.
The relevant data for the model of~\cite{Taylor:2015xtz} has been serialized in the file \verb|1511-03209-resolved.mrdi|~\cite{bmtMaRDIFiles}, allowing the following computation to execute within a few seconds on most personal computers:
\begin{minted}[bgcolor=bg,linenos,breaklines]{julia}
julia> g4_amb_candidates = chosen_g4_flux_gens(t_res, completeness_check = false);

julia> length(g4_amb_candidates)
629
\end{minted}
Despite the simplicity of our approach, it effectively identifies 480 basis elements of $H^{2, 2}(X_\Sigma, \mathbb{Q})$ that restrict trivially to $\widehat{Y}_4$. This enables us to parametrize the $G_4$-flux candidates using only the remaining 629 generators.

\subsubsection{\texorpdfstring{$G_4$}{G4}-Flux Families}

As discussed in \cref{subsec:Fluxes-Quantization} and \cref{subsec:Fluxes-FurtherProperties}, the computation of intersection numbers is fundamental in identifying $G_4$-flux candidates with prescribed properties. However, performing these computations via Gr{\"o}bner basis techniques in the cohomology ring of the toric ambient space is infeasible for the model of~\cite{Taylor:2015xtz}. To circumvent this limitation, we introduce an alternative algorithm that efficiently determines intersection numbers, even in this intricate geometric setting. The details of this algorithm are presented in \cref{subsec:AlgoSophisticatedIntersection}.

Using this algorithm, we systematically identify families of $G_4$-fluxes. Our initial computations, performed from scratch, require several hours, as the algorithm identifies and processes millions of intersection numbers. However, these results have now been fully serialized, drastically reducing computation time. While our advanced intersection product computations were essential for obtaining the initial results, serialization reduces the runtime of flux family computations to mere seconds on most personal computers:
\begin{minted}[bgcolor=bg,linenos,breaklines]{julia}
julia> fg_quant = special_flux_family(t_res, completeness_check = false)
Family of G4 fluxes:
  - Elementary quantization checks: satisfied
  - Transversality checks: satisfied
  - Non-abelian gauge group: breaking pattern not analyzed

julia> size(matrix_integral(fg_quant))
(629, 224)

julia> size(matrix_rational(fg_quant))
(629, 127)
\end{minted}
This result shows that the candidate $G_4$-fluxes passing elementary quantization and transversality checks are parametrized by $\mathbb{Z}^{224} \times \mathbb{Q}^{127}$. Similarly, we can invoke non-abelian gauge group breaking:
\begin{minted}[bgcolor=bg,linenos,breaklines]{julia}
julia> fg_quant_no_break = special_flux_family(t_res, not_breaking = true, completeness_check = false)
Family of G4 fluxes:
  - Elementary quantization checks: satisfied
  - Transversality checks: satisfied
  - Non-abelian gauge group: unbroken

julia> size(matrix_integral(fg_quant_no_break))
(629, 1)

julia> size(matrix_rational(fg_quant_no_break))
(629, 127)
\end{minted}

For most F-theory applications, these computations are typically refined further. Specifically, one would seek to isolate the subset of well-quantized $G_4$-fluxes that either preserve the non-abelian gauge group or break it to a desired subgroup. Additionally, these fluxes would be subjected to the D3-tadpole constraint. We encourage interested readers to utilize our tools to perform these computations for the given geometry. A systematic study of these aspects is reserved for future work.

\section{Summary and Outlook} \label{sec:Outlook}

In this paper, we have introduced \FTheoryTools, an innovative computational module integrated into the \texttt{OSCAR} computer algebra system~\cite{OSCAR, OscarBook}. This tool is designed to tackle some of the most pressing computational challenges in F-theory research. Key functionalities include the enumeration of vertical $G_4$-fluxes, the ability to perform blowups on arbitrary loci, and an ever-growing database of F-theory geometries adhering to FAIR principles~\cite{fair}. These features aim to simplify the workflow of researchers, making complex F-theory computations readily accessible.

As a proof of concept, we have demonstrated the application of \FTheoryTools to a highly complex geometry known as \emph{the F-theory geometry with most flux vacua}~\cite{Taylor:2015xtz}. In this setup, we model vertical $G_4$-fluxes using cohomology classes of the toric ambient space. \FTheoryTools can then identify candidate well-quantized and transverse $G_4$-fluxes. The space of identified candidate fluxes is isomorphic to $\mathbb{Z}^{224} \times \mathbb{Q}^{127}$. The exact details are encoded in the arithmetic results that we make available at~\cite{bmtMaRDIFiles}.

We envision several exciting directions for the development of \FTheoryTools in the future. These include both enhancements to existing functionalities and the introduction of new features, as outlined below.

One of the immediate goals is to improve the user interface for $G_4$-flux computations, enabling more intuitive and efficient workflows. In addition, we aim to extend the tool's capability to analyze $G_4$-fluxes that partially or entirely preserve the gauge group. These features would provide deeper insights into flux vacua and their physical implications.

As noted earlier, the current database of F-theory geometries is far from complete. A key priority is to expand this repository by incorporating additional constructions. Furthermore, support for larger model classes, such as complete intersection models and generalized CICYs~\cite{Anderson:2015iia}, should be provided.
These developments would enhance the tool's utility, and may also help to foster more effective collaboration within the F-theory community.

In future updates, we plan to include explicit functionality for the analysis of higher-codimension loci, and to include support for chiral and vector-like spectra computations, leveraging recent advancements in the field such as~\cite{Bies:2018uzw, Bies:2023jqg, Li:2023dya}. The roadmap also includes the incorporation of automated resolution approaches for certain geometries.

To promote reproducibility and transparency, we wish to investigate how \FTheoryTools could benefit from established proof systems such as \texttt{LEAN} \cite{deMoura2015lean, leanprover}. This could help prevent unnecessary computations and allow us to verify at least some results, making \FTheoryTools more robust.

To support the community, we plan to extend the existing tutorial and enhance the documentation, offering detailed guidance on advanced use cases. This will empower users to fully leverage the capabilities of \FTheoryTools in their research. Let us use this opportunity to direct the reader to the documentation\footnote{This link will cease to function once \FTheoryTools joins the \OSCAR core. The interested reader should then visit \href{https://docs.oscar-system.org/stable/}{\small{\mbox{https://docs.oscar-system.org/stable/}}} and use the search function to identify the relevant sections in the documentation.}:
\begin{center}
    \href{https://docs.oscar-system.org/stable/Experimental/FTheoryTools/introduction/}{https://docs.oscar-system.org/stable/Experimental/FTheoryTools/introduction/}.
\end{center}
A notebook containing all code snippets from this article, along with a general-purpose introductory tutorial to \FTheoryTools, is available at:
\begin{center}
    \href{https://github.com/HereAround/MartinsOscarTutorials.jl/blob/master/FTheoryToolsPaper.ipynb}{https://github.com/HereAround/MartinsOscarTutorials.jl/blob/master/FTheoryToolsPaper.ipynb},
    \href{https://www.oscar-system.org/tutorials/FTheoryTools/}{https://www.oscar-system.org/tutorials/FTheoryTools/}.
\end{center}

The development of \FTheoryTools is an ongoing process, with the ultimate goal of becoming a robust and reliable open-source software solution for F-theory research. By addressing the challenges outlined above, we hope to make this tool an indispensable resource for the community, bridging the gap between mathematical abstraction and practical computation.

The utility of \texttt{LiteratureModels} increases with the number of available constructions, yet extracting precise model data from publications remains a time-consuming task. We warmly encourage researchers to contribute models of their own or those of particular interest. Contributions---such as PDFs of relevant papers, JSON files with extracted parameters, or MaRDI files as described in \cref{mardi_sec}---are all welcome and can be sent to us via email (see the author information on the first page of this article).

\paragraph{Acknowledgements}
The authors express their gratitude and appreciation for the support provided by the \OSCAR team, led by Simon Brandhorst, Claus Fieker, Tommy Hofmann, Max Horn, Michael Joswig, and Wolfram Decker.

We express special thanks to Antony Della Veccia for his support and work on the \texttt{MaRDI} file format; to Matthias Zach for creating the scheme framework within \OSCAR upon which the non-toric blowup technology rests; and Matthias Zach and Lars Kastner for their work on the toric--scheme interface. We also thank James Halverson, Nikhil Raghuram, and Mohab Safey El Din for helpful discussions.

The authors are thankful for the support offered by the \emph{TU-Nachwuchsring}. This work was supported by the SFB-TRR 195 \emph{Symbolic Tools in Mathematics and their Application} of the German Research Foundation (DFG). Martin Bies acknowledges financial support from the \emph{Forschungsinitiative des Landes Rheinland-Pfalz} through the project \emph{SymbTools -- Symbolic Tools in Mathematics and their Application}. Andrew P. Turner acknowledges funding from \emph{DOE (HEP) Award DE-SC0013528} and NSF grant \emph{PHY-2014086}.

\newpage

\appendix

\section{Specialized \texorpdfstring{$G_4$}{G4}-Flux Algorithms}
\label{sec:AppendixAlgorithms}

Throughout this section, we continue to make all assumptions outlined in \cref{sec:EnumerationOfG4Fluxes}.

\subsection{Finding a Generating Set}
\label{subsec:AlgoDetailsGenFinding}

Given a simplicial, complete toric variety $X_\Sigma$, our objective is to compute a basis for $H^{2, 2}(X_\Sigma, \mathbb{Q}) = H^4(X_\Sigma, \mathbb{Q})$ (cf.\ Theorem~9.3.2 of~\cite{CLS:2011}). Theorem~12.4.1 of~\cite{CLS:2011} establishes an isomorphism
\begin{equation}
    H^4(X_\Sigma, \mathbb{Q}) \cong R_{\mathbb{Q}}(\Sigma)_2\,,
\end{equation}
where $R_{\mathbb{Q}}(\Sigma)_2$ denotes the space of all degree-2 monomials (with respect to the standard grading) in the cohomology ring $R_{\mathbb{Q}}(\Sigma)$. This isomorphism allows us to construct a basis in which elements of $H^{2, 2}(X_\Sigma, \mathbb{Q})$ can be expanded with rational coefficients. The cohomology ring of $X_\Sigma$ is given as the quotient
\begin{equation}
    R_{\mathbb{Q}}(\Sigma) = \mathbb{Q}[x_1, \dots, x_r] / \left( I_\text{lin} + I_\text{SR} \right)\,,
\end{equation}
where $r$ denotes the number of rays in $\Sigma$, while $I_\text{lin}$ and $I_\text{SR}$ denote the ideal of linear relations and the Stanley--Reisner ideal, respectively.

The generic \OSCAR algorithm \verb|basis_of_h4| determines a basis of $R_{\mathbb{Q}}(\Sigma)_2$ using Gr{\"o}bner basis computations. However, in the context of F-theory models such as~\cite{Taylor:2015xtz}, this approach is computationally prohibitive. Moreover, the special structure of $R_{\mathbb{Q}}(\Sigma)$ permits the design of a significantly more efficient method. Consequently, \FTheoryTools employs a simple specialized algorithm to construct a generating set of vertical $G_4$-flux candidates.

This specialized algorithm computes a basis for $H^{2, 2}(X_\Sigma, \mathbb{Q})$ by exploiting the structure of the toric variety. The fundamental idea is to construct the basis directly from toric divisors by forming quadratic monomials of their associated cohomology classes. The implementation proceeds as follows:
\begin{enumerate}
    \item Extract the linear relations among toric divisors and process them to determine an independent set of generators for $H^{1, 1}(X_\Sigma, \mathbb{Q})$.
    \item Form quadratic monomials from these generators, omitting elements that belong to the Stanley--Reisner ideal, as they vanish identically.
    \item Eliminate any remaining linear dependencies among these quadratic elements to yield a basis for $H^{2, 2}(X_\Sigma, \mathbb{Q})$.
\end{enumerate}
This method efficiently identifies a basis of $H^{2, 2}(X_\Sigma, \mathbb{Q})$.

\subsection{Restricting the Generating Set to \texorpdfstring{$\widehat{Y}_4$}{Y4}} 
\label{subsec:AlgoDetailsRestriction}

Given a simplicial, complete toric variety $X_\Sigma$ and a basis of $H^{2, 2}(X_\Sigma, \mathbb{Q})$, we seek to restrict these basis elements to the cohomology of $\widehat{Y}_4$. To this end, we consider an F-theory geometry $\widehat{Y}_4$ given by a hypersurface,
\begin{equation}
    \widehat{Y}_4 = \mathbb{V}(p) \subset X_\Sigma.
\end{equation}
Elements of $H^{2, 2}_\text{vert}(\widehat{Y}_4, \mathbb{Q})$ are then obtained by restricting the basis of $H^{2, 2}(X_\Sigma, \mathbb{Q})$ to $\widehat{Y}_4$.

We focus on determining whether a given cohomology class on $X_\Sigma$---represented by the algebraic cycle $\mathbb{V}(x_a, x_b)$, where $x_a$ and $x_b$ are homogeneous coordinates in the toric ambient space $X_\Sigma$---restricts trivially to the hypersurface $\mathbb{V}(p)$. Ideally, if the algebraic cycles are in general position, their intersection is transverse, and it suffices to check whether their closed intersection is empty. However, practical challenges arise.

First, while toric geometry allows us to identify a rationally equivalent cycle to a given algebraic cycle (using the linear relations of $X_\Sigma$), verifying general position is nontrivial. A standard approach involves computing the vanishing locus of all $k$ equations defining the involved algebraic cycles. If the resulting intersection variety forms a codimension-$k$ complete intersection, the cycles are considered to be in general position. However, this reasoning is somewhat circular: we adjust algebraic cycles to achieve general position in order to compute their set-theoretic intersection, yet we rely on this intersection to confirm that the cycles were in general position in the first place.

Second, even if general position is assumed, one might attempt to use the \emph{Toric Weak Nullstellensatz} (Proposition~5.2.6 in~\cite{CLS:2011}) to determine whether the set-theoretic intersection is empty. However, this approach requires ideal membership checks, which are computationally infeasible for geometries as complex as those in~\cite{Taylor:2015xtz}, given the excessive size of the associated cohomology ring.

To circumvent these difficulties, we avoid both general position verification and the \emph{Toric Weak Nullstellensatz}. Instead, for a cohomology class represented by $\mathbb{V}(x_a, x_b)$ and a hypersurface $\mathbb{V}(p)$, we employ the following procedure:
\begin{enumerate}
    \item Evaluate the hypersurface equation at $x_a = x_b = 0$, i.e., compute $s = p(x_a = 0, x_b = 0)$.\footnote{If $x_a = x_b$, evaluate $p(x_a = 0)$ instead. Although the algebraic cycle $\mathbb{V}(x_a, x_a)$ could be replaced by a rationally equivalent linear combination of algebraic cycles, we omit this step to maintain consistency with our decision to avoid the general position question.}
    \item The Stanley--Reisner ideal of $X_\Sigma$ prevents certain variables from vanishing simultaneously with $x_a$ and $x_b$. Set these restricted variables to $1$ in $s$ by utilizing the scaling relations of $X_\Sigma$, yielding a modified polynomial $\tilde{s}$.
    \item If $\tilde{s}$ simplifies to a nonzero constant, then the cohomology class represented by $\mathbb{V}(x_a, x_b)$ restricts trivially to $\mathbb{V}(p)$. Such cohomology classes are discarded, while all others are retained.
\end{enumerate}

It is important to emphasize that this approach may sometimes fail to correctly determine whether a cohomology class vanishes upon restriction to the hypersurface. Our checks are sufficient but not necessary: we do not test whether two cohomology classes become identical upon restriction to $\mathbb{V}(p)$. Consequently, the computed generating set of $G_4$-fluxes may be larger than strictly required---some elements may vanish upon restriction to $\mathbb{V}(p)$ in a way that our algorithm does not detect, while others may secretly be equivalent upon restriction.

\subsection{A Specialized Approach to Intersection Products}
\label{subsec:AlgoSophisticatedIntersection}

To identify families of $G_4$-flux candidates with specific properties, it is essential to compute intersection numbers. Specifically, we aim to evaluate integrals of the form:
\begin{equation}
    \int_{X_\Sigma}{g \wedge \PD{H} \wedge \PD{D_i} \wedge \PD{D_j}}\,.
\end{equation}
Here, $g$ denotes a cohomology class corresponding to one of the $G_4$-flux generators obtained via the command \verb|chosen_g4_flux_gens|, incorporating the steps outlined in \cref{subsec:AlgoDetailsGenFinding} and \cref{subsec:AlgoDetailsRestriction}. We assume that the cohomology class $g$ is represented by an algebraic cycle $\mathbb{V}(x_a, x_b)$ for appropriate indices $a$ and $b$. Similarly, $\PD{D_i}$ and $\PD{D_j}$ represent cohomology classes Poincar{\'e} dual to toric divisors, respectively corresponding to the algebraic cycles $\mathbb{V}(x_i)$ and $\mathbb{V}(x_j)$. Finally, $\PD{H}$ denotes the cohomology class Poincar{\'e} dual to the hypersurface $\mathbb{V}(p) = \widehat{Y}_4$.

A naive approach to computing such intersection products would be:
\begin{equation}
    \int_{X_\Sigma}{g \wedge \PD{H} \wedge \PD{D_i} \wedge \PD{D_j}} \stackrel{?}{=} \Bigl|\left\{ x_i = x_j = x_a = x_b = p = 0 \right\}\Bigr| \, .
\end{equation}
The issue is that the intersection may not be transverse. For instance, in the case where $i = j = a = b$, intersection theory necessitates moving the algebraic cycles into general position. In toric geometry, this is achieved by leveraging the linear relations of the toric space, which allow for the replacement of one algebraic cycle with a rational linear combination of others.

For the toric ambient space $X_\Sigma$ of~\cite{Taylor:2015xtz}, five linear relations exist, computable in \OSCAR as the generators of \verb|ideal_of_linear_relations|. By iteratively applying these relations, we successfully avoid self-intersections in all relevant computations for~\cite{Taylor:2015xtz}.

Formally, after (potentially multiple) applications of the linear relations, our algorithm determines
\begin{equation}
    \int_{X_\Sigma}{g \wedge \PD{H} \wedge \PD{D_i} \wedge \PD{D_j}} = \sum_{k = 1}^N \lambda_k \Bigl|\left\{x_{a_k} = x_{b_k} = x_{c_k} = x_{d_k} = p = 0\right\} \Bigr| \,.
\end{equation}
The coefficients $\lambda_k \in \mathbb{Q}$, the number of terms $N$, and the indices $a_k, b_k, c_k, d_k$---each distinct for a given $k \in \{1, \dotsc, N\}$---depend on the specific linear relations applied. Notably, these data are not unique.

Ideally, for pairwise distinct indices $a_k, b_k, c_k, d_k$, the following variety would be a reduced complete intersection:
\begin{equation}
    \left\{ x_{a_k} = x_{b_k} = x_{c_k} = x_{d_k} = p = 0 \right\} \, .
\end{equation}
However, this is not always the case. In practice, $p(x_{a_k} = x_{b_k} = x_{c_k} = x_{d_k} = 0)$ may be a power of another polynomial, e.g., $p(x_{a_k} = x_{b_k} = x_{c_k} = x_{d_k} = 0 ) = \tilde{p}^2$. Alternatively, the hypersurface polynomial may vanish identically, i.e., $p(x_{a_k} = x_{b_k} = x_{c_k} = x_{d_k} = 0) = 0$. Our algorithm mitigates these issues by iteratively applying linear relations, moving the algebraic cycles into sufficiently general position until $p(x_{a_k} = x_{b_k} = x_{c_k} = x_{d_k} = 0)$ satisfies one of a few predefined cases from which the intersection number can be readily determined.

Two technical aspects of our algorithm warrant mention. First, for a complete, simplicial toric variety, the linear relations define a vector space of algebraic cycles rationally equivalent to zero. When replacing an algebraic cycle defined by four distinct homogeneous coordinates and the hypersurface equation, one of the four coordinates is randomly selected for replacement via a linear relation. Though this process is nondeterministic, our experience indicates that this approach greatly reduces the computation time for large models by comparison with the standard algorithm employed in \OSCAR. Second, the specific algorithm used for the geometry of~\cite{Taylor:2015xtz} relies on a predefined set of well-defined intersection numbers relevant for that geometry. Applying this algorithm to different spaces may require additional hard-coded cases.

Finally, we enumerate the currently implemented hard-coded cases explicitly. The sufficiency of these cases depends on the specific model geometry under consideration. In its current form, the specialized intersection number algorithm is verified to work with the model of~\cite{Taylor:2015xtz} and its resolution described in \cref{subsubsec:RecapOfBigModel}, as well as to all QSM geometries~\cite{CHLLT19}, as discussed in \cref{subsec:ComputingG4s}. While it may apply to other geometries, we generally expect that additional hard-coded cases would need to be incorporated.

To formalize our discussion, let $p$ denote the strict transform of the Tate polynomial of~\cite{Taylor:2015xtz}. We evaluate $p$ as four toric coordinates, say $x_i$, vanish. By utilizing the Stanley--Reisner ideal of the toric ambient space $X_\Sigma$, certain homogeneous coordinates $y_j$ must not vanish simultaneously with the $x_i$. The $\mathbb{Z}^n$-grading of the Cox ring of $X_\Sigma$ allows us to rescale all such non-vanishing homogeneous coordinates $y_j$ to $1$. After using these scaling relations, the condition $0 = p(x_i = 0, y_j = 1)$ is then expressed in terms of the remaining coordinates $z_k$ of $X_\Sigma$.

In fact, there will always be only two remaining coordinates after this process, $z_1, z_2$. This is because the set of rays corresponding to $x_{a_k}, x_{b_k}, x_{c_k}, x_{d_k}$ (assuming they intersect) form a four-dimensional cone, and there must be exactly two five-dimensional (maximal) cones in the complete, smooth five-dimensional ambient space that have this four-dimensional cone as a facet. Each of these two five-dimensional cones has one additional ray, and these rays correspond to the coordinates $z_1, z_2$. The only remaining Stanley--Reisner relation (and thus the only remaining scaling relation) that has not been used is between these remaining coordinates $z_1, z_2$. We then only need to hard-code cases for polynomials in two variables $z_1, z_2$ that cannot simultaneously vanish and have one scaling relation relating them. The only simplification we choose to make that is not fully justified is to limit these hard-coded cases to low-degree polynomials of particular forms. This is the area in which the algorithm may need to be generalized to apply to other geometries.

The hard-coded cases we use here are as follows:
\begin{enumerate}
    \item If $p(x_i = 0, y_j = 1) = q_1 \cdot z_1 + q_2 \cdot z_2$, where $q_1, q_2 \in \mathbb{Q} \setminus \{0\}$, and $z_1$, $z_2$ must not vanish simultaneously by virtue of the Stanley--Reisner ideal, with $z_1, z_2$ subject to the (unused) scaling relation 
    \begin{equation}
        \left[z_1 \colon z_2\right] \sim \left[ \lambda z_1 \colon \lambda z_2 \right]\,, \quad \lambda \in \mathbb{C} \setminus \{ 0 \}\,,
    \end{equation}
    then the equation $0 = p(x_i = 0, y_j = 1)$ has exactly one solution, namely $z_1 = 1, z_2 = -\frac{q_1}{q_2}$, which is equivalent via the scaling relation to $z_1 = -\frac{q_2}{q_1}, z_2 = 1$.
    
    \item If $p(x_i = 0, y_j = 1) = q \cdot z_1$, where $q \in \mathbb{Q} \setminus \{ 0 \}$, and $z_1$, $z_2$ must not vanish simultaneously by virtue of the Stanley--Reisner ideal, with $z_1, z_2$ subject to the (unused) scaling relation
    \begin{equation}
        \left[ z_1 \colon z_2 \right] \sim \left[ \lambda^{n_1} z_1 \colon \lambda^{n_2} z_2 \right]\,, \quad \lambda \in \mathbb{C} \setminus \{ 0 \}\,, \quad n_1, n_2 \in \mathbb{Z}\,, \quad n_2 \ne 0\,,
    \end{equation}
    then the equation $0 = p(x_i = 0, y_j = 1)$ has exactly one solution, namely $z_1 = 0$ and $z_2 = 1$.
    
    \item If $p(x_i = 0, y_j = 1) = q \cdot z_1 \cdot z_2$, where $q \in \mathbb{Q} \setminus \{ 0 \}$, and $z_1$, $z_2$ must not vanish simultaneously by virtue of the Stanley--Reisner ideal, with $z_1, z_2$ subject to the (unused) scaling relation
    \begin{equation}
        \left[z_1 \colon z_2\right] \sim \left[ \lambda^{n_1} z_1 \colon \lambda^{n_2} z_2 \right]\,, \quad \lambda \in \mathbb{C} \setminus \{ 0 \}\,, \quad n_1, n_2 \in \mathbb{Z}\,, \quad n_1, n_2 \ne 0\,,
    \end{equation}
    then the equation $0 = p(x_i = 0, y_j = 1)$ has exactly two solutions: $(z_1, z_2) = (1,0)$ and $(z_1, z_2) = (0,1)$.

    \item If $p(x_i = 0, y_j = 1) = q_1 \cdot z_1^2 + q_2 \cdot z_1 \cdot z_2 + q_3 \cdot z_2^2$, where $q_1, q_2, q_3 \in \mathbb{Q} \setminus \{ 0 \}$, and $z_1$, $z_2$ must not vanish simultaneously by virtue of the Stanley--Reisner ideal, with $z_1, z_2$ subject to the (unused) scaling relation
    \begin{equation}
        \left[z_1 \colon z_2\right] \sim \left[ \lambda^{n_1} z_1 \colon \lambda^{n_2} z_2 \right]\,, \quad \lambda \in \mathbb{C} \setminus \{ 0 \}\,, \quad n_1, n_2 \in \mathbb{Z}\,,
    \end{equation}
    where $n_1, n_2$ are not both zero, then the equation $0 = p(x_i = 0, y_j = 1)$ has exactly two solutions, counted with multiplicity.
\end{enumerate}

\bibliographystyle{JHEP}
\bibliography{references}

\end{document}